\documentclass[aps,prl,longbibliography,twocolumn,superscriptaddress, nobibnotes, 10pt, floatfix]{revtex4-2}
\usepackage[utf8]{inputenc}
\usepackage[T1]{fontenc}
\usepackage{amsmath,amssymb}
\usepackage{times}
\usepackage{xcolor}
\usepackage{physics}
\usepackage[normalem]{ulem}
\usepackage{xspace}
\usepackage{float}
\usepackage{comment}
\usepackage[colorlinks=true,
            linkcolor=teal,
            urlcolor=teal,
            citecolor=teal,
            filecolor=magenta]{hyperref}
\usepackage{mathtools}
\usepackage{graphicx}
\usepackage{dcolumn}
\usepackage{bm}
\usepackage{orcidlink}

\usepackage{amsmath}
\usepackage{bm}
\usepackage{commath}
\newcommand{\RR}{\mathbb{R}}
\newcommand{\II}{\mathbb{I}}

\newcommand{\calD}{\mathcal{D}}
\newcommand{\vect}[1]{\bm{#1}}

\usepackage[normalem]{ulem}

\begin{document}


\title{Universality of superdiffusion in simple random graphs}
\author{Mrinal Sarkar\,\orcidlink{0000-0003-3112-6530}}
\email{Corresponding author; email: sarkar@thphys.uni-heidelberg.de}
\affiliation{Institut für Theoretische Physik, Universität Heidelberg, 69120 Heidelberg, Germany 
}
\author{Nicolò Defenu\,\orcidlink{0000-0002-3401-3665}}%
\affiliation{Institut für Theoretische Physik, ETH Zürich, 8093 Zürich, Switzerland
}%
\author{Tilman Enss\,\orcidlink{0000-0002-5334-2448}}%
\affiliation{Institut für Theoretische Physik, Universität Heidelberg, 69120 Heidelberg, Germany 
}%


\begin{abstract}
Random walks with long-range jumps can drive superdiffusive transport, replacing ordinary diffusion with an effective long-range kinetic operator. Such superdiffusive kinetics is also central to critical phenomena, notably the self-avoiding walk with long-range jump statistics, or L\'evy-SAW. This work investigates how the critical behavior is affected when the long-range connectivity itself becomes random. We study self-avoiding walks (SAWs) on a one-dimensional long-range random ring graph, where bonds are independently generated with Bernoulli probability $\sim|i-j|^{-(1+\sigma)}$. We term this walk Sparse-SAW. The same random bonds are responsible for both long-range superdiffusive transport and quenched disorder, with both simultaneously controlled by the single parameter $\sigma$, placing the problem beyond the conventional Harris and Weinrib-Halperin frameworks. Through large-scale Monte Carlo simulations and a Gaussian-truncated field theory, we show that Sparse-SAW belongs to the same universality class as the clean superdiffusive L\'evy-SAW. The random bonds generate short-range uncorrelated and long-range correlated mass disorder while simultaneously producing the long-range kinetic operator. Under coarse-graining, the latter dominates, restoring the clean critical behavior. Our study suggests that the full non-Gaussian Bernoulli statistics may lead to disorder physics beyond the conventional theory of quenched disorder, while establishing random graphs as an efficient platform for extracting the critical exponents of the clean superdiffusive L\'evy-SAW universality class.
\end{abstract}

\maketitle

\section{Significance}
Random graphs provide a natural description of many real-world systems, from brain wiring to social interactions. Their irregular connectivity is itself a form of disorder. Whether such randomness changes the collective behavior is a fundamental question. We show that, surprisingly, a self-avoiding walk on such a random graph with long-range links exhibits the same universal critical behavior as a clean system with long-range motion. We uncover that the random connections not only introduce disorder but also determine how the walk proceeds, with this long-range motion governing the large-scale physics. Such a graph disorder behaves fundamentally differently from a conventional one, providing a new perspective on critical phenomena in complex systems.

\section{Introduction}
Diffusion is among the most fundamental transport processes in nature, yet on graphs, networks, and other disordered structures it frequently gives way to anomalous behavior. In particular, long-range connectivity can give rise to superdiffusive transport, in which long-distance pathways generate an effective long-range kinetic description~\cite{metzler2000random,riascos2014fractional,juhasz2008superdiffusion}, as also realized experimentally\,\cite{Barthelemy2008}. Superdiffusion has also recently emerged as an important theme in quantum many-body systems, including long-range interacting spin chains, dephasing-driven transport, hydrodynamic regimes, and disordered quantum circuits~\cite{catalano2023anomalous,anand2026robustness,capizzi2025hydrodynamics,mcculloch2026kpz}. This broad interest highlights the relevance of understanding the origin and robustness of superdiffusive behavior across different physical settings. In the context of critical phenomena, such long-range kinetics also underlies a class of critical statistical systems, notably self-avoiding walks with long-range steps, whose large-scale behavior can be described by stable L\'evy processes~\cite{heydenreich2009long,chen2010asymptotic}. A natural question is therefore how the critical behavior associated with superdiffusive kinetics is affected when the underlying geometry is itself disordered.

Random graphs with long-range bonds offer a natural playground to probe this question. A walker exploring such a graph inherits an LR kinetic operator $\sim|q|^\sigma$ directly from the bond statistics, in place of the ordinary diffusive $q^2$~\cite{metzler2000random,riascos2014fractional,juhasz2008superdiffusion}. Indeed, the simple, non-interacting random walks on long-range percolation clusters have been shown to exhibit scaling limits of stable L\'evy processes~\cite{crawford2013simple,berger2024scaling}. However, little is known about how self-avoidance interacts with graph disorder.

We investigate the self-avoiding walk (SAW) on the recently introduced simple (no self-loops or multi-edges) random graph: the one-dimensional long-range random ring (1DLR3)\,\cite{millan2021complex, sarkar2024universality}. We call the resulting walk Sparse-SAW. The critical behavior of the clean self-avoiding L\'evy flight (L\'evy-SAW) on homogeneous lattices is well understood. Unlike the L\'evy-SAW, the Sparse-SAW walker moves on a quenched random graph. In a 1DLR3 graph, a bond between any pair of nodes $(i,j)$ is added \emph{independently} with probability $p_{ij}=|i-j|^{-(1+\sigma)}$~\cite{millan2021complex}. The bonds thus follow a Bernoulli distribution, and the same bonds are responsible for long-distance transport while also acting as a source of quenched disorder.

Quenched disorder in the form of random bonds, random fields, impurities, or lattice defects can alter the critical properties near a second-order phase transition~\cite{vojta2019disorder,dahlberg2025spin}. The Harris criterion offers a foundational criterion for weak, spatially short-range (SR) uncorrelated disorder: the clean critical point remains stable against the disorder if the correlation-length exponent of the clean system satisfies $\nu>2/d$, with $d$ being the spatial dimension~\cite{harris1974effect}. Otherwise, disorder drives the system to a new random fixed point. Weinrib and Halperin extended the Harris criterion to disorder with algebraically decaying correlations $\sim r^{-a}$~\cite{weinrib1983critical}. For $a>d$, these long-range (LR) correlations are irrelevant and the standard Harris criterion applies. For $a<d$, the disorder is irrelevant only if $\nu>2/a$; otherwise, a new disordered fixed point emerges. This underscores the role of disorder correlations in critical phenomena\,\cite{liu1999kosterlitz, rieger1999random, schrenk2013percolation, ibrahim2014enhanced, mendes2019localization, modak2020many}.

However, recent studies of criticality in such disordered graphs have challenged this conventional understanding, finding that several universality classes including XY, and self-avoiding walk retain the critical behavior of their clean, long-range counterparts despite the presence of disorder~\cite{berganza2013critical,cescatti2019analysis,sarkar2024universality, bighin2024universal}. In our 1DLR3 graphs, a single parameter $\sigma$ sets both the kinetic and disorder sectors simultaneously. This problem therefore lies fundamentally outside the conventional Harris and Weinrib--Halperin (H-WH) framework, in which disorder perturbs a pre-existing clean theory. We ask two questions. First, do the conventional H-WH disorder-irrelevance criteria apply when the disorder is not the standard additive Gaussian field assumed by Harris and WH, but is instead non-Gaussian (Bernoulli) in nature and simultaneously tied to the kinetic sector? Second, does the resulting universality class remain that of the clean superdiffusive L\'evy-SAW?

A similar random graph, where bonds occur with power-law probability and random sign, appears in diluted long-range Ising spin glasses~\cite{leuzzi2008dilute}. The non-Gaussian Bernoulli dilution was argued to yield the same universality class as that of fully connected model with Gaussian couplings~\cite{leuzzi2008dilute}. This shows the relevance of our graphs beyond this immediate context. Moreover, they provide a computationally efficient platform for studying long-range critical phenomena, while raising the broader question of when non-Gaussian correlated disorder can remain irrelevant at criticality.

We demonstrate, through large-scale Monte Carlo simulations, that the critical exponents of Sparse-SAW agree with those of the clean, superdiffusive L\'evy-SAW throughout the region of interest. This is enabled by a new algorithm for generating an effectively infinite graph, together with a branching-SAW algorithm developed in our earlier work~\cite{sarkar2025long}. By mapping the SAW onto the $n\to0$ limit of the $O(n)$ model, a Gaussian-truncated field theory reveals that the random bonds generate both SR uncorrelated and LR correlated random-mass disorder, while simultaneously producing the LR kinetic term itself. However, it is the LR kinetic operator that dominates under coarse-graining, restoring the clean universality class. Moreover, contrary to the Weinrib--Halperin case, where more strongly correlated disorder is generically more dangerous, our graph disorder maintains an approximately constant irrelevance margin already at the Gaussian level, highlighting its fundamentally different nature.

\section*{Results}
\paragraph*{Numerical results: Critical exponents--}To establish our main results, we perform large-scale Monte Carlo (MC) simulations of SAWs on the 1DLR3 graph. Two algorithmic advances make this possible. First, to minimize the finite-size effects, we develop an efficient way of generating an infinite 1DLR3 graph on the fly during the random walk, rather than pre-constructing a finite graph, see Method Sec.~\ref{sec:methods-graph}. The idea is as follows: whenever a walker visits a node for the first time, the node is \emph{explored}, meaning its links are drawn as if the graph were infinite, following the prescribed distribution, while excluding previously visited nodes. This strategy ensures the walker effectively explores an infinite graph. Since the links are generated only while \emph{exploring} a node, the computational cost scales linearly as $\mathcal{O}(\langle \kappa \rangle L)$. Second, standard simulations on graphs yield typical walks of lengths too small to reach the scaling regime. The available efficient algorithms such as the pivot method are not applicable to random graph geometries. We overcome this using the branching SAW algorithm of Ref.\,\cite{sarkar2025long}: at each node, the walker branches out making $z (> 1)$ attempts for the next step: the first step is always performed, an additional step is attempted with probability $z - 1$. Convergence requires $z < z_c$.  As the susceptibility $\chi(z) \sim (z_c -z)^{-\gamma} \quad \text{for} \quad z \nearrow z_c$, tuning $z \to z_c(\sigma)$ from below allows arbitrarily long walks to be sampled. The walks are truncated at a maximum length $N$. The branching factor satisfies $z_c(\sigma \to 0) = 1$ and $z_c(\sigma \to \infty) = 2$, and is determined numerically for each $\sigma$. The SAW critical 
exponents $\nu_{\rm LR}$ and $\gamma$ are defined by
\begin{equation}
\langle \log R_N \rangle \sim \nu_{\rm LR} \log N, \qquad 
c_N \sim \mu_{\text{SAW}}^N N^{\gamma - 1},
\label{eq:exponent_def}
\end{equation}
where $R_N$ is the end-to-end distance, $c_N$ is the number of  $N$-step SAWs, and $\mu_{\text{SAW}}$ is the connective constant. The logarithmic average in the first relation is used in place of $\langle R_N^2\rangle$ 
since the heavy-tailed jump distribution renders the mean squared displacement ill-defined\,\cite{grassberger1985critical, sarkar2025long}.

We now proceed to investigate the behavior of the critical exponents $\nu_{\rm LR}$ and $\gamma$ in this paper. In our MC simulation, for each $\sigma$, at least $10^9$ ($10^7$) independent samples are generated for shorter (longer) walks for walks upto $N=2 \times 10^3$ steps.

\begin{figure}[t]
    \centering
    \includegraphics[width=0.95\linewidth]{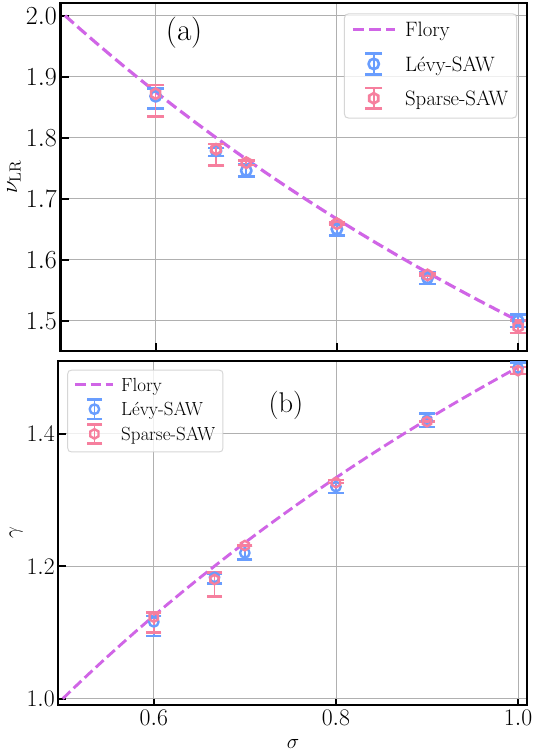}
    \caption{Critical exponents $\nu_{\rm LR}$ and $\gamma$ at various $\sigma$ in the LR regime of clean theory ($0.5 < \sigma < 1$). Sparse-SAW estimates (red hexagons) agree with the clean L\'evy-SAW results (blue circles, Ref.\,\cite{sarkar2025long}), within statistical uncertainties. The dashed violet line corresponds to Flory prediction, and empty square represents the mean-field value.}
    \label{fig:gamma_vs_sigma_LR}
\end{figure}

\paragraph*{Long-range behavior--}We first turn our attention to the LR regime of the clean system, $1/2 <\sigma<1$. We compute $\nu_{\rm LR}$ and $\gamma$ as a function of $\sigma$ across this regime, see Fig.~\ref{fig:gamma_vs_sigma_LR}. Our analysis follows the approach adopted in Ref.\,\cite{sarkar2025long}, where finite-walk-length exponents $\nu_{\rm LR}(N)$ and $\gamma(N)$ are first computed from the scaling of $\langle \log R_N \rangle$ and $c_N$ respectively, and then asymptotic estimates are obtained by incorporating correction to scaling. Our MC estimates (red hexagons) are close to but consistently deviate from the clean LR Flory prediction (dashed violet line). For intermediate $\sigma$, we observe clear scaling regimes, and the corrections to scaling are consistent with the theoretical prediction that they are universal\,\cite{sokal1994monte} and depend on the distance from the LR-SR boundary. The downward deviation compared to Flory is not fully conclusive from Fig.~\ref{fig:gamma_vs_sigma_LR} as one approaches the LR–SR crossover boundary, which is attributed to strong finite-size effects. Crucially, the same small deviation from Flory is also present in the clean self-avoiding L\'evy-flight estimates (blue circles: L\'evy-SAW), reported in Ref.\,\cite{sarkar2025long}, with which our Sparse-SAW results agree within error bars throughout. This indicates that the departure from Flory reflects a genuine feature of the clean universality class itself and Flory is an approximate, rather than exact, prediction, which Sparse-SAW inherits unchanged. This agreement with the clean model serves as direct evidence that the quenched graph disorder is irrelevant for the critical behavior of Sparse-SAW. 

\begin{figure}[t]
    \centering    \includegraphics[width=0.9\linewidth]{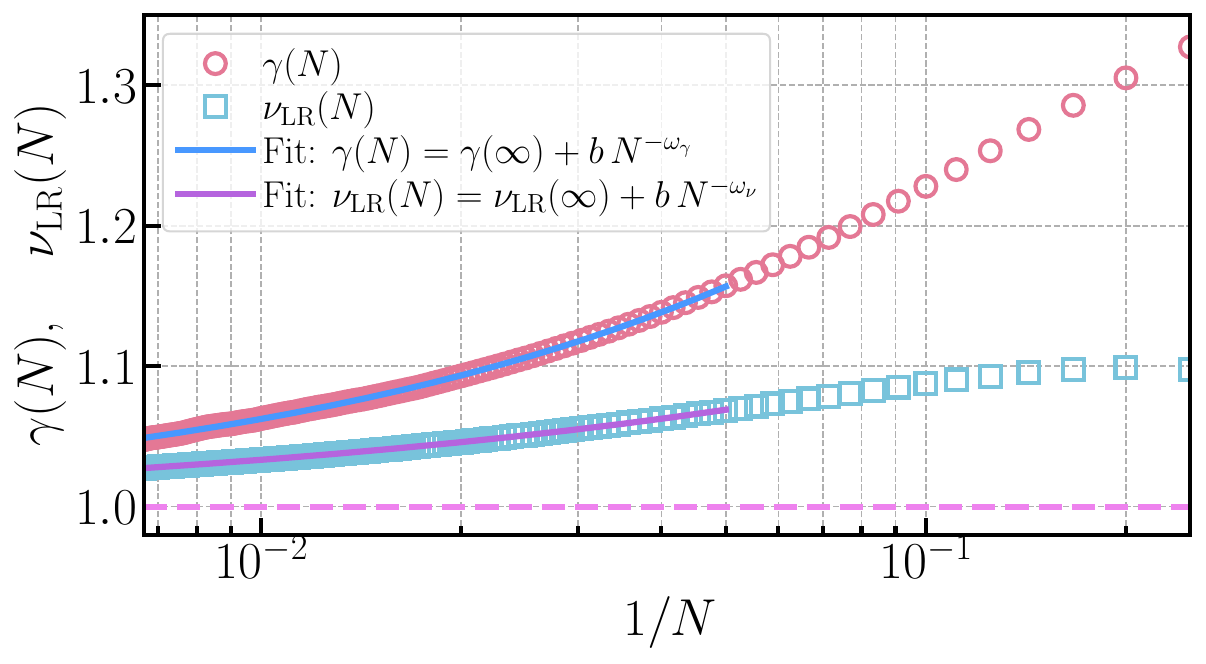}
    \caption{Scaling of finite-$N$ exponents $\nu_{\rm {LR}}(N)$ and $\gamma(N)$ in the SR regime of the clean kinetic sector at $\sigma = 1.5$. They reach asymptotically to clean SR fixed point $\nu_{\rm {LR}}(\infty) = \gamma (\infty) = 1 $ with correction-to-scaling exponents $\omega_{\nu} = \omega_{\gamma} \approx \sigma - \sigma^{*}$, consistent with the field-theoretic predictions.
    }
    \label{fig:gamma_scaling_SR}
\end{figure}

\paragraph*{Short-range behavior--} For $\sigma \geq 1$, we analyze the scaling of $\nu_{\rm LR}(N)$ and $\gamma(N)$ with walk length $N$. In this regime, rather than focusing on the asymptotic values of $\nu_{\rm LR}$ and $\gamma$, it is instructive to examine how they approach this limit. We now explicitly demonstrate convergence to the clean SR universality class at $\sigma=1.5$ in Fig.~\ref{fig:gamma_scaling_SR}. Such a choice within $\sigma \in [1,2]$ minimizes the boundary effects and associated numerical uncertainties. Indeed, fitting the finite-walk-length data with the field theory expectations $\nu_{\rm LR}(N) = \nu_{\rm LR}(\infty) + b\,N^{-\omega_{\nu}}$ and $\gamma(N) = \gamma(\infty) + b\,N^{-\omega_{\gamma}}$ with the universal correction-to-scaling exponent $\omega_{\nu} = \sigma - \sigma^* = \omega_{\gamma}$ yields $\gamma (\infty) = 0.980^{+0.018}_{-0.008}, \nu(\infty) = 1.001^{+0.002}_{-0.007} \qquad \text{and} \quad \omega_{\gamma} = 0.49^{+0.05}_{-0.03} \approx \sigma - \sigma^{*} \approx 0.51^{+0.02}_{-0.04} = \omega_{\nu}$, which are consistent with the clean SR fixed-point $\nu(\infty) = \gamma (\infty) = 1$.These results are stable against variation of the window sizes following the procedure in \,\cite{sarkar2025long}.  This provides a direct numerical confirmation that LR correlation in bond disorder is irrelevant for critical behavior in SAWs. Note that such a precise verification is made possible by our branching algorithm and effectively infinite graph generation, which substantially improves the statistics for long walks.

\subsection*{Field theory}
This section develops a field-theoretic description of our model to explain, at the microscopic level, why Sparse-SAW shares the same universality class as the clean, superdiffusive L\'evy-SAW.

\paragraph*{Emergent correlated mass disorder:}
To probe the effect of graph disorder on criticality, we start with an $n$-vector $O(n)$ model on the 1DLR3 graph, with the Hamiltonian
\begin{equation}
    \beta H = - \frac{1}{2} \sum_{i \neq j} J_{ij} \vect{S}_i \cdot \vect{S}_j,  \qquad 
    \vect{S}_i \in \RR^n,
\label{eq:On-Hamiltonian}
\end{equation}
where the normalization $|\vect{S}_i|^2 = n$ follows de Gennes formulation\,\cite{de1972exponents}. The couplings $J_{ij}$ are independent random variables drawn from the following Bernoulli distribution
\begin{equation}
P(J_{ij}) = p_{ij} \, \delta(J_{ij} - 1) + (1 - p_{ij}) \, \delta(J_{ij}),
\label{eq:bond_dist}
\end{equation}
so that all bonds which are present with probability $p_{ij}$ are of unit strength. We will finally focus on the $n \to 0$ limit, which maps onto self-avoiding walks (SAWs)~\cite{de1972exponents}.

To access the field-theoretic description, we map the $O(n)$ model on the graph to a continuous $|\vect{\phi}|^4$ field theory via a Hubbard-Stratonovich transformation\,\cite{ nishimori2010elements, machado2010local}. Introducing an auxiliary field $\vect{\phi}_i \in \RR^n$, this yields for the partition function
\begin{equation}
Z =  \mathcal{N} \int \calD \vect{\phi}\, \exp\left(-\frac{1}{2} \vect{\phi}^T \cdot \vect{K}^{-1} \cdot \vect{\phi}  + \sum_j U (\vect{\phi}_j) \right)
\end{equation}
where $\mathcal{N}$ is a normalization constant. The single-site potential, defined by,
\begin{equation}
    e^{ U (\vect{\phi}_i)} =  \int d\vect{S}_j \, \delta(|\vect{S}_j|^2 - n) e^{- \vect{\phi}_i^T \cdot \vect{S}_j}
    \label{eq:U_phi}
\end{equation}
evaluates to $\exp{(U(\vect{\rho}))} = f(n) \, (\sqrt{\rho})^{1- \frac{n}{2}} \, I_{\frac{n}{2} -1} (\sqrt{n \rho})$,
where $f(n) = n^{\frac{n}{4} - \frac{1}{2}} \, (2\pi)^{\frac{n}{2}}$ and $I_{\nu} (x)$ is the modified Bessel function of the first kind. A Taylor expansion of the $U(\rho)$ around $\rho$ or $\phi = 0$ generates the mass and the interaction terms of the field theory, $U(\vect{\phi}) = \text{const.} + (1/2) |\vect{\phi}|^2 - (1/4(n+2)) |\vect{\phi}|^4 + \mathcal{O} (|\vect{\phi}|^6)$. The $nN \times nN$ symmetric coupling matrix is $\vect{K}_{ij} = \vect{J}_{ij} + 2\mu \II \, \delta_{ij}$, where $\II$ is an identity matrix and $\mu$ is a constant chosen to make the matrix positive definite. Further contribution to the mass comes from the diagonal elements of the $\vect{K}^{-1}$. This yields, up to $O(J^2)$ in the sparse-graph limit,$ (\vect{K}^{-1})_{ii} =  (1/2\mu) \left[1 + {z_i}/(4\mu^2) \right]$, where $z_i := \sum_{k \neq i} J_{ik}$ is the degree (number of neighbors) of the $i$-th node on the graph.

The effective mass term on a given graph realization is
\begin{equation}
 \frac{1}{2} \sum_i \left( -1 + \frac{1}{2\mu} + \frac{z_i}{8\mu^3} + \dots \right) |\vect{\phi}_i|^2.
\end{equation}
The degree $z_i$ is a random variable that fluctuates from site to site, implying that \emph{random bond disorder generates random mass disorder}. 
Writing $z_i = \bar{z} + \delta z_i$, $\bar{z}$ being the average degree, the question arises, how is this mass disorder $\delta z_i$ correlated? One can easily show that the disorder has both a SR uncorrelated part and a LR correlated part\,\cite{SM}.
\begin{equation}
\boxed{
\overline{\delta z_i \delta z_j} = 
\begin{cases}
\displaystyle \sum_{k \neq i} p_{ik}(1-p_{ik}) \approx \bar{z}, & i = j, \\[1em]
p_{ij}(1-p_{ij}) \approx \dfrac{1}{|i-j|^{1+\sigma}}, & i \neq j.
\end{cases}
}
\label{eq:mass-corr}
\end{equation}
This could be intuitively understood as follows:
(1) \textbf{SR uncorrelated disorder ($i = j$)}: The variance of $\delta z_i$ is the sum of variances of all independent bonds emanating from the site $i$. Each bond contributes $p_{ik}(1-p_{ik})$, which in the sparse graph limit yields $\bar{z}$.
(2) \textbf{LR correlations ($i \neq j$)}: Since each bond is independent, the only way $\delta z_i$ and $\delta z_j$ can be correlated is if there is a direct bond between sites $i$ and $j$. The correlation is exactly the variance of that single bond, which is $p_{ij}(1-p_{ij}) \approx p_{ij}$. Thus, the correlation decays as $1/|i-j|^{1+\sigma}$, the same power law as the bond probability.

\paragraph*{Gaussian truncation and disorder averaging via the replica trick:}
We now proceed to obtain the disorder-averaged replicated action. We truncate the emergent mass disorder at the level of its first two cumulants and treat as an effectively Gaussian random field coupled linearly to the local energy density with zero mean and variance given by Eq.\,\eqref{eq:mass-corr}. The LR bonds being source of quenched disorder, we employ the replica trick, $\overline{\ln Z} = \lim_{m \to 0} (\overline{Z^m} - 1)/m$, to perform the average over graph disorder\,\cite{dotsenko2005introduction}. In the continuum limit, this yields the effective replicated action\,\cite{SM}
\begin{align}
S_{\text{eff}}[\vect{\phi}] = & \frac{1}{2} \sum_{A=1}^m \int dx \, dy \, \frac{\vect{\phi}_A(x) \cdot \vect{\phi}_A(y)}{|x-y|^{1+\sigma}} \nonumber \\
&+ r\, \sum_{A=1}^m \int dx \, |\vect{\phi}_A(x)|^2
+ u \, \sum_{A=1}^m \int dx \, |\vect{\phi}_A(x)|^4 \nonumber \\
&- \frac{v_{\rm SR}}{2} \sum_{A,B}{}' \int dx \,|\vect{\phi}_A(x)|^2|\vect{\phi}_B(x)|^2 \nonumber \\
&- \frac{v_{\rm LR}}{2} \sum_{A,B}{}' \int dx\, dy \, \frac{|\vect{\phi}_A(x)|^2|\vect{\phi}_B(y)|^2}{|x-y|^{1+\sigma}}.
\label{eq:action_graph}
\end{align} 
where $A, B$ are replica indices and $\sum'_{A,B}$ restricts the sum to $A\neq B$. The effective mass term is $r= \frac{1}{2} \left ( -1 + \frac{1}{2 \mu + \bar{z}} + \frac{\bar{z}}{8 \mu^3} \right )$, and $u= \frac{1}{4(n+2)} - \frac{v_{\rm SR}}{2}$.
The same-replica ($A=B$) disorder contraction also generates a further nonlocal single-replica self-interaction of the same order as $v_{\rm LR}$, which does not affect disorder-relevance conclusions below, see \,\cite{SM}. The disorder-induced inter-replica interaction is a short-range Harris-type term, with coefficient $v_{\rm SR}$, together with a long-range Weinrib--Halperin-type term, with coefficient $v_{\rm LR}$.

\paragraph*{RG relevance of disorder on Sparse-SAW:} Before proceeding, let us briefly recapitulate the disorder relevance criteia for SAW\,\cite{chakrabarti1981statistics, kim1983renormalisation, meir1989self, grassberger1993recursive,  barat1995statistics, kim1987self}. For SR uncorrelated disorder, the disorder does not introduce a new direction in the coupling space, rather merely renormalizes the pure quartic local coupling. So Harris criterion does not apply; clean and SR correlated disordered SAWs belong to the same universality class\,\cite{kim1983renormalisation}. For LR correlated disorder, one obtains a single disorder-relevance criterion: disorder is irrelevant if $\nu > 2/a$, irrespective of $a<d$ or $a>d$\,\cite{kim1987self}.

We now ask the relevance of disorder in the Gaussian-truncated replicated action, Eq.~\eqref{eq:action_graph}, for Sparse-SAW specifically. A crucial observation is that the bond disorder simultaneously generates both the mass disorder and the kinetic term itself, so there is no pre-existing clean fixed point to perturb around. The field's scaling dimension is fixed at tree level by the LR kinetic term alone $[\vect{\phi}]_{\rm LR} = (d-\sigma)/2$. The relevance of each disorder sector follows directly from the engineering dimension of its associated operator. Our region of interest is $d/2<\sigma<\sigma^*$ and $\sigma^*<\sigma<2$, where $\sigma^*$ marks the LR to SR crossover boundary of the clean kinetic term (discussed later). The lower bound $\sigma=d/2$ marks the mean-field to long-range boundary, set by the scaling dimension of the self-interaction, $[u]=2\sigma-d$. In the both region of our interest, the theory flows to an interacting fixed point. 

For $d/2 < \sigma < \sigma^*$, the scaling dimension of the local energy-density operator at the clean fixed point $[\vect{\phi}^2 (x)] = d - 1/\nu_{\rm LR}$\,\cite{cardy1996scaling}, which determines the scaling dimension of the LR disorder-induced interactions $[v_{\rm LR}] = {2}/{\nu_{\rm LR}} - a$ with $a=1+\sigma$. As argued above, SR disorder has no effect on the Sparse-SAW either. Thus the disorder-irrelevance criterion for Sparse-SAW reads as\begin{equation}
\nu_{\rm LR} > \frac{2}{1 + \sigma}.
\label{eq:dis_rel_sparse_SAW}
\end{equation}

The conclusion for $\sigma>d$, however, requires care. Here, $[\vect{\phi}]_{\text{LR}} <0$ signals that the LR fixed point ceases to be the correct infrared description, and the kinetic sector itself simultaneously undergoes a crossover to SR theory. Following Sak's criterion, the crossover occurs at $\sigma^* = 2- \eta_{\text{SR}}$, with $\eta_{\text{SR}}$ the anomalous dimension in SR limit\,\cite{sak1973recursion, brezin2014crossover, defenu2015fixed, behan2017long, behan2017scaling, benedetti2025one, sarkar2025long, li20264epsilon}. For $\sigma^* < \sigma < 2$, the SR kinetic term generated under coarse-graining becomes relevant and governs the infrared fixed point. Following Ref.\,\cite{kim1987self}, the disorder-irrelevance criterion in Eq.\eqref{eq:dis_rel_sparse_SAW} continues to apply, now evaluated at the SR fixed point, yielding $\nu_{\rm SR} > {2}/{(1+\sigma)}$. Note that for $d \geq 2$, $\eta_{\text{SR}} >0$ implies $\sigma^* < d$, so the crossover to SR behavior occurs before $\sigma$ reaches $d$. In $d=1$, $\eta_{\text{SR}}$ is defined only in the $n \to 0$ limit, where $\eta_{\text{SR}} = 1$ and thus $\sigma^* = d =1$.

From our recent work\,\cite{sarkar2025long}, the measured exponent $\nu_{\rm LR}$ of the clean L\'evy-SAW remains close to, but deviates slightly from, the Flory prediction $\nu_{\rm Flory}=3/(1+\sigma)$\,\cite{sarkar2025long,grassberger1985critical,halley1985node,de1979scaling} throughout the LR regime. Thus in this regime, the disorder-irrelevance criterion for Sparse-SAW, Eq.\eqref{eq:dis_rel_sparse_SAW}, is comfortably satisfied, maintaining an approximately constant irrelevance margin because $\nu_{\rm Flory}/[2/(1+\sigma)] = 3/2$ is independent of $\sigma$. On the SR side, using $\nu_{\rm SR}=1$, the criterion becomes $\sigma>1=\sigma^*$, satisfied throughout $\sigma^*<\sigma<2$, and marginal at $\sigma=\sigma^*$ itself. This lets us establish our main message: the correlated graph disorder is harmless for Sparse-SAW regardless of regime, governed by clean field theory.

\section*{Discussion}
To summarize, we have shown that the random bonds on the simple 1DLR3 graph leaves the critical behavior of Sparse-SAW unchanged from that of the clean superdiffusive L\'evy-SAW. On the numerical side, we have developed a new algorithm to generate an effectively infinite graph. Together with a branching-SAW algorithm, we generate sufficiently long walks near criticality to access the true asymptotic scaling regime, overcoming the strong corrections to scaling that had obscured it in earlier studies, and obtain the asymptotic estimates of the exponents~\cite{sarkar2025long}. To validate, a Gaussian-truncated field theory shows that the random bonds that generate this correlated mass disorder also generate the LR kinetic term itself, placing the system beyond the standard Harris or Weinrib--Halperin (H-WH) paradigm, in which a pre-existing clean fixed point is perturbed by disorder independently\,\cite{harris1974effect,weinrib1983critical}. Furthermore, in sharp contrast with WH scenario, an increase in disorder correlations make disorder more dangerous, here the disorder maintains an approximately constant irrelevance margin throughout the long-range regime.

Two remarks are in order. First, the Bernoulli bond distribution generates non-Gaussian mass disorder where all higher-order cumulants decay $\sim |i-j|^{-(1+\sigma)}$ at the leading order) and are expected to produce additional disorder channels beyond the Gaussian approximation. Averaging over the full Bernoulli distribution may therefore lead to behavior beyond the Gaussian-truncated theory. Already at the Gaussian level, disorder exhibits a trend opposite to the Weinrib--Halperin scenario, suggesting that higher-order cumulants could yield qualitatively new effects. Second, the present disorder-irrelevance result is specific to the $n\to0$ replica limit describing SAW, where the short-range disorder channel is irrelevant. For general $O(n\ge1)$ models on the same graph, the short-range disorder is expected to be relevant. Understanding this general $O(n\ge1)$ together with the full non-Gaussian Bernoulli statistics may uncover new disorder-driven universality classes and constitutes a promising direction for future work.

Our results, together with our recent demonstration of localization and quantum-chaotic behavior on the same random graph\,\cite{sarkar2026emergent}, suggest that Bernoulli graph disorder should be viewed as a distinct class of disorder in critical phenomena rather than as a straightforward extension of conventional quenched disorder. This disorder irrelevance demonstrates that these graphs can serve as an efficient numerical platform to extract clean superdiffusive critical exponents. More broadly, advances in programmable quantum simulators and photonic networks that enable the engineering of graph-based geometries open promising avenues for exploring critical phenomena on random graphs experimentally\,\cite{qiang2021implementing, nguyen2023quantum}.

\section{Methods}
\label{sec:methods-graph}
\small{
\paragraph{Numerical implementation: Generating the graph efficiently}
This section provides an efficient way to generate an infinite 1DLR3 graph. To this end, consider an infinite linear graph with nodes $i$, with $i = -\infty, \dots, -1, 0, 1, \dots, \infty$. As mentioned, any two nodes $(i, j)$ of length $l=|i-j|$ on the graph are connected by a link with probability
\begin{equation}
    p_{ij} = p_l = |i - j|^{-a}, \qquad \text{with}\qquad a=1+\sigma.
    \label{eqn:app-prob}
\end{equation}
According to this distribution, nearest neighbors are always connected and provide a backbone. Note that all links are independent, and they are set or not set irrespective of the presence of other links.

The idea to generate the graph is to draw a list of links for every node, which scales linearly as $O(\langle \kappa \rangle L)$. The average degree $\langle \kappa \rangle$ depends on the value of $\sigma$. To generate the list of links for a given node $i$, one first constructs the cumulative probability distribution function for drawing the next link. The probability that the next link has length $l = 2$ is given by $p_2$ (skipping the backbone $l = 1$), but if $l = 2$ is not chosen, then the probability for length $l = 3$ is $(1 - p_2) p_3$, and so on:
\begin{align}
    P_2 &= p_2, \\
    P_3 &= (1 - p_2) p_3, \\
    P_4 &= (1 - p_2)(1 - p_3) p_4, \dots
\end{align}

This resembles a geometric distribution but with different $p_i$. The cumulative distribution function (CDF) that a link up to length $l$ is chosen next is given by:
\begin{equation}
    c_l = \sum_{i=2}^{l} P_i = 1 - \prod_{i=2}^{l} (1 - p_i), \quad c_1 = 0.
\end{equation}

Even for an infinite number of possible links, $c_\infty < 1$. The next link is drawn as follows:
\begin{enumerate}
    \item Set the lower bound of the CDF to $c_{\min} = 0$.
    \item Draw a random number uniformly as $r \in [c_{\min}, 1)$.
    \item If $r \geq c_\infty$, no further link is drawn (list is complete).
    \item Otherwise, the next link has length $l$ such that $c_{l-1} \leq r < c_l$.
    \item Set the new lower bound $c_{\min} = c_l$ and repeat from step 2.
\end{enumerate}

For a power-law distribution, the CDF $c_l$ is given explicitly as:
\begin{equation}
    c_l = 1 - \exp \left[ - \sum_{k=1}^{\infty} \frac{1}{k} \left( \zeta(ka, 2) - \zeta(k a, l+1) \right) \right]
\end{equation}
where $\zeta(s, q)$ is the generalized zeta function,
\begin{equation}
    \zeta(s, q) = \sum_{i=q}^{\infty} i^{-s}.
\end{equation}

For universal properties, large-$l$ limit of the distribution (\ref{eqn:app-prob}) plays the dominant role. We thus approximate:
\begin{equation}
    \sum_{k=1}^{\infty} \frac{1}{k} \zeta(ka, q) \approx \frac{(q - 1/2)^{-(a - 1)}}{a - 1}.
\end{equation}

Then, the approximate CDF becomes
\begin{align}
    c_\infty &\approx 1 - \exp \left[ -\frac{(3/2)^{-\sigma}}{\sigma} \right], \\
    c_l &\approx 1 - (1 - c_\infty) \exp \left[ \frac{(l + 1/2)^{-\sigma}}{\sigma} \right].
\end{align}
The inverse function is
\begin{equation}
    l = \left( \sigma \ln \frac{1 - c_l}{1 - c_\infty} \right)^{-1/\sigma} - \frac{1}{2}.
\end{equation}

One may generate a finite graph by applying a similar algorithm. For a finite graph on a ring of length $L$ one must ensure $c_L = 1$ to prevent links back to the same node.

In numerics, SAW starts at position $i = 0$. At each step, links are drawn from position $i$ using the above algorithm. The walker decides to move to one of its neighbors by choosing a link randomly. However, the move is accepted if the new position has never been visited before. The walk moves to the corresponding new position, and explore the node: links are drawn following the algorithm, excluding previously visited nodes. On the other hand, if this position has already been visited, the walk terminates.
}

\section{Data availability}{All MC simulation code including graph-generation algorithm and raw-data supporting our findings will be made available to readers in a public repository (Github) upon publication.}

\section{Acknowledgments}{This project is supported by the Deutsche Forschungsgemeinschaft (DFG, German Research Foundation) under Germany’s Excellence Strategy EXC 2181/1-390900948 (the Heidelberg STRUCTURES Excellence Cluster). M.S. also acknowledges support by the state of Baden-Württemberg through the bwHPC.}

\bibliography{References_mrinal}

\begin{thebibliography}{55}%
\makeatletter
\providecommand \@ifxundefined [1]{%
 \@ifx{#1\undefined}
}%
\providecommand \@ifnum [1]{%
 \ifnum #1\expandafter \@firstoftwo
 \else \expandafter \@secondoftwo
 \fi
}%
\providecommand \@ifx [1]{%
 \ifx #1\expandafter \@firstoftwo
 \else \expandafter \@secondoftwo
 \fi
}%
\providecommand \natexlab [1]{#1}%
\providecommand \enquote  [1]{``#1''}%
\providecommand \bibnamefont  [1]{#1}%
\providecommand \bibfnamefont [1]{#1}%
\providecommand \citenamefont [1]{#1}%
\providecommand \href@noop [0]{\@secondoftwo}%
\providecommand \href [0]{\begingroup \@sanitize@url \@href}%
\providecommand \@href[1]{\@@startlink{#1}\@@href}%
\providecommand \@@href[1]{\endgroup#1\@@endlink}%
\providecommand \@sanitize@url [0]{\catcode `\\12\catcode `\$12\catcode
  `\&12\catcode `\#12\catcode `\^12\catcode `\_12\catcode `\%12\relax}%
\providecommand \@@startlink[1]{}%
\providecommand \@@endlink[0]{}%
\providecommand \url  [0]{\begingroup\@sanitize@url \@url }%
\providecommand \@url [1]{\endgroup\@href {#1}{\urlprefix }}%
\providecommand \urlprefix  [0]{URL }%
\providecommand \Eprint [0]{\href }%
\providecommand \doibase [0]{https://doi.org/}%
\providecommand \selectlanguage [0]{\@gobble}%
\providecommand \bibinfo  [0]{\@secondoftwo}%
\providecommand \bibfield  [0]{\@secondoftwo}%
\providecommand \translation [1]{[#1]}%
\providecommand \BibitemOpen [0]{}%
\providecommand \bibitemStop [0]{}%
\providecommand \bibitemNoStop [0]{.\EOS\space}%
\providecommand \EOS [0]{\spacefactor3000\relax}%
\providecommand \BibitemShut  [1]{\csname bibitem#1\endcsname}%
\let\auto@bib@innerbib\@empty
\bibitem [{\citenamefont {Metzler}\ and\ \citenamefont
  {Klafter}(2000)}]{metzler2000random}%
  \BibitemOpen
  \bibfield  {author} {\bibinfo {author} {\bibfnamefont {R.}~\bibnamefont
  {Metzler}}\ and\ \bibinfo {author} {\bibfnamefont {J.}~\bibnamefont
  {Klafter}},\ }\bibfield  {title} {\bibinfo {title} {The random walk's guide
  to anomalous diffusion: a fractional dynamics approach},\ }\href
  {https://doi.org/https://doi.org/10.1016/S0370-1573(00)00070-3} {\bibfield
  {journal} {\bibinfo  {journal} {Physics Reports}\ }\textbf {\bibinfo {volume}
  {339}},\ \bibinfo {pages} {1} (\bibinfo {year} {2000})}\BibitemShut {NoStop}%
\bibitem [{\citenamefont {Riascos}\ and\ \citenamefont
  {Mateos}(2014)}]{riascos2014fractional}%
  \BibitemOpen
  \bibfield  {author} {\bibinfo {author} {\bibfnamefont {A.~P.}\ \bibnamefont
  {Riascos}}\ and\ \bibinfo {author} {\bibfnamefont {J.~L.}\ \bibnamefont
  {Mateos}},\ }\bibfield  {title} {\bibinfo {title} {Fractional dynamics on
  networks: Emergence of anomalous diffusion and l\'evy flights},\ }\href
  {https://doi.org/10.1103/PhysRevE.90.032809} {\bibfield  {journal} {\bibinfo
  {journal} {Phys. Rev. E}\ }\textbf {\bibinfo {volume} {90}},\ \bibinfo
  {pages} {032809} (\bibinfo {year} {2014})}\BibitemShut {NoStop}%
\bibitem [{\citenamefont {Juh\'asz}(2008)}]{juhasz2008superdiffusion}%
  \BibitemOpen
  \bibfield  {author} {\bibinfo {author} {\bibfnamefont {R.}~\bibnamefont
  {Juh\'asz}},\ }\bibfield  {title} {\bibinfo {title} {Superdiffusion in a
  class of networks with marginal long-range connections},\ }\href
  {https://doi.org/10.1103/PhysRevE.78.066106} {\bibfield  {journal} {\bibinfo
  {journal} {Phys. Rev. E}\ }\textbf {\bibinfo {volume} {78}},\ \bibinfo
  {pages} {066106} (\bibinfo {year} {2008})}\BibitemShut {NoStop}%
\bibitem [{\citenamefont {Barthelemy}\ \emph {et~al.}(2008)\citenamefont
  {Barthelemy}, \citenamefont {Bertolotti},\ and\ \citenamefont
  {Wiersma}}]{Barthelemy2008}%
  \BibitemOpen
  \bibfield  {author} {\bibinfo {author} {\bibfnamefont {P.}~\bibnamefont
  {Barthelemy}}, \bibinfo {author} {\bibfnamefont {J.}~\bibnamefont
  {Bertolotti}},\ and\ \bibinfo {author} {\bibfnamefont {D.~A.}\ \bibnamefont
  {Wiersma}},\ }\bibfield  {title} {\bibinfo {title} {A {L}\'evy flight for
  light},\ }\href {https://doi.org/10.1038/nature06948} {\bibfield  {journal}
  {\bibinfo  {journal} {Nature}\ }\textbf {\bibinfo {volume} {453}},\ \bibinfo
  {pages} {495} (\bibinfo {year} {2008})}\BibitemShut {NoStop}%
\bibitem [{\citenamefont {Catalano}\ \emph {et~al.}(2023)\citenamefont
  {Catalano}, \citenamefont {Mattiotti}, \citenamefont {Dubail}, \citenamefont
  {Hagenm\"uller}, \citenamefont {Prosen}, \citenamefont {Franchini},\ and\
  \citenamefont {Pupillo}}]{catalano2023anomalous}%
  \BibitemOpen
  \bibfield  {author} {\bibinfo {author} {\bibfnamefont {A.~G.}\ \bibnamefont
  {Catalano}}, \bibinfo {author} {\bibfnamefont {F.}~\bibnamefont {Mattiotti}},
  \bibinfo {author} {\bibfnamefont {J.}~\bibnamefont {Dubail}}, \bibinfo
  {author} {\bibfnamefont {D.}~\bibnamefont {Hagenm\"uller}}, \bibinfo {author}
  {\bibfnamefont {T.}~\bibnamefont {Prosen}}, \bibinfo {author} {\bibfnamefont
  {F.}~\bibnamefont {Franchini}},\ and\ \bibinfo {author} {\bibfnamefont
  {G.}~\bibnamefont {Pupillo}},\ }\bibfield  {title} {\bibinfo {title}
  {Anomalous diffusion in the long-range haken-strobl-reineker model},\ }\href
  {https://doi.org/10.1103/PhysRevLett.131.053401} {\bibfield  {journal}
  {\bibinfo  {journal} {Phys. Rev. Lett.}\ }\textbf {\bibinfo {volume} {131}},\
  \bibinfo {pages} {053401} (\bibinfo {year} {2023})}\BibitemShut {NoStop}%
\bibitem [{\citenamefont {Anand}\ \emph {et~al.}(2026)\citenamefont {Anand},
  \citenamefont {Kemp}, \citenamefont {Wei}, \citenamefont {White},
  \citenamefont {Zaletel},\ and\ \citenamefont {Yao}}]{anand2026robustness}%
  \BibitemOpen
  \bibfield  {author} {\bibinfo {author} {\bibfnamefont {S.}~\bibnamefont
  {Anand}}, \bibinfo {author} {\bibfnamefont {J.}~\bibnamefont {Kemp}},
  \bibinfo {author} {\bibfnamefont {J.}~\bibnamefont {Wei}}, \bibinfo {author}
  {\bibfnamefont {C.~D.}\ \bibnamefont {White}}, \bibinfo {author}
  {\bibfnamefont {M.~P.}\ \bibnamefont {Zaletel}},\ and\ \bibinfo {author}
  {\bibfnamefont {N.~Y.}\ \bibnamefont {Yao}},\ }\href
  {https://arxiv.org/abs/2602.15933} {\bibinfo {title} {Robustness of
  kardar-parisi-zhang-like transport in long-range interacting quantum spin
  chains}} (\bibinfo {year} {2026}),\ \Eprint
  {https://arxiv.org/abs/2602.15933} {arXiv:2602.15933} \BibitemShut {NoStop}%
\bibitem [{\citenamefont {Capizzi}\ \emph {et~al.}(2025)\citenamefont
  {Capizzi}, \citenamefont {Wang}, \citenamefont {Xu}, \citenamefont {Mazza},\
  and\ \citenamefont {Poletti}}]{capizzi2025hydrodynamics}%
  \BibitemOpen
  \bibfield  {author} {\bibinfo {author} {\bibfnamefont {L.}~\bibnamefont
  {Capizzi}}, \bibinfo {author} {\bibfnamefont {J.}~\bibnamefont {Wang}},
  \bibinfo {author} {\bibfnamefont {X.}~\bibnamefont {Xu}}, \bibinfo {author}
  {\bibfnamefont {L.}~\bibnamefont {Mazza}},\ and\ \bibinfo {author}
  {\bibfnamefont {D.}~\bibnamefont {Poletti}},\ }\bibfield  {title} {\bibinfo
  {title} {Hydrodynamics and the eigenstate thermalization hypothesis},\ }\href
  {https://doi.org/10.1103/PhysRevX.15.011059} {\bibfield  {journal} {\bibinfo
  {journal} {Phys. Rev. X}\ }\textbf {\bibinfo {volume} {15}},\ \bibinfo
  {pages} {011059} (\bibinfo {year} {2025})}\BibitemShut {NoStop}%
\bibitem [{\citenamefont {McCulloch}(2026)}]{mcculloch2026kpz}%
  \BibitemOpen
  \bibfield  {author} {\bibinfo {author} {\bibfnamefont {E.}~\bibnamefont
  {McCulloch}},\ }\href {https://doi.org/https://arxiv.org/abs/2608.06459}
  {\bibinfo {title} {{KPZ} superdiffusion of local correlators in diffusive
  random quantum circuits}} (\bibinfo {year} {2026}),\ \Eprint
  {https://arxiv.org/abs/2608.06459} {arXiv:2608.06459} \BibitemShut {NoStop}%
\bibitem [{\citenamefont {Heydenreich}(2011)}]{heydenreich2009long}%
  \BibitemOpen
  \bibfield  {author} {\bibinfo {author} {\bibfnamefont {M.}~\bibnamefont
  {Heydenreich}},\ }\bibfield  {title} {\bibinfo {title} {Long-range
  self-avoiding walk converges to $\alpha$-stable processes},\ }\href
  {https://doi.org/10.1214/09-AIHP350} {\bibfield  {journal} {\bibinfo
  {journal} {Annales de l'I.H.P. Probabilités et statistiques}\ }\textbf
  {\bibinfo {volume} {47}},\ \bibinfo {pages} {20} (\bibinfo {year}
  {2011})}\BibitemShut {NoStop}%
\bibitem [{\citenamefont {Chen}\ and\ \citenamefont
  {Sakai}(2011)}]{chen2010asymptotic}%
  \BibitemOpen
  \bibfield  {author} {\bibinfo {author} {\bibfnamefont {L.-C.}\ \bibnamefont
  {Chen}}\ and\ \bibinfo {author} {\bibfnamefont {A.}~\bibnamefont {Sakai}},\
  }\bibfield  {title} {\bibinfo {title} {{Asymptotic behavior of the gyration
  radius for long-range self-avoiding walk and long-range oriented
  percolation}},\ }\href {https://doi.org/10.1214/10-AOP557} {\bibfield
  {journal} {\bibinfo  {journal} {The Annals of Probability}\ }\textbf
  {\bibinfo {volume} {39}},\ \bibinfo {pages} {507 } (\bibinfo {year}
  {2011})}\BibitemShut {NoStop}%
\bibitem [{\citenamefont {Crawford}\ and\ \citenamefont
  {Sly}(2013)}]{crawford2013simple}%
  \BibitemOpen
  \bibfield  {author} {\bibinfo {author} {\bibfnamefont {N.}~\bibnamefont
  {Crawford}}\ and\ \bibinfo {author} {\bibfnamefont {A.}~\bibnamefont {Sly}},\
  }\bibfield  {title} {\bibinfo {title} {Simple random walk on long-range
  percolation clusters ii: Scaling limits},\ }\href
  {https://doi.org/10.1214/12-AOP774} {\bibfield  {journal} {\bibinfo
  {journal} {Ann. Probab.}\ }\textbf {\bibinfo {volume} {41}},\ \bibinfo
  {pages} {445} (\bibinfo {year} {2013})}\BibitemShut {NoStop}%
\bibitem [{\citenamefont {Berger}\ and\ \citenamefont
  {Tokushige}(2024)}]{berger2024scaling}%
  \BibitemOpen
  \bibfield  {author} {\bibinfo {author} {\bibfnamefont {N.}~\bibnamefont
  {Berger}}\ and\ \bibinfo {author} {\bibfnamefont {Y.}~\bibnamefont
  {Tokushige}},\ }\href {https://arxiv.org/abs/2403.18532} {\bibinfo {title}
  {Scaling limits for random walks on long range percolation clusters}}
  (\bibinfo {year} {2024}),\ \Eprint {https://arxiv.org/abs/2403.18532}
  {arXiv:2403.18532} \BibitemShut {NoStop}%
\bibitem [{\citenamefont {Mill{\'a}n}\ \emph {et~al.}(2021)\citenamefont
  {Mill{\'a}n}, \citenamefont {Gori}, \citenamefont {Battiston}, \citenamefont
  {Enss},\ and\ \citenamefont {Defenu}}]{millan2021complex}%
  \BibitemOpen
  \bibfield  {author} {\bibinfo {author} {\bibfnamefont {A.~P.}\ \bibnamefont
  {Mill{\'a}n}}, \bibinfo {author} {\bibfnamefont {G.}~\bibnamefont {Gori}},
  \bibinfo {author} {\bibfnamefont {F.}~\bibnamefont {Battiston}}, \bibinfo
  {author} {\bibfnamefont {T.}~\bibnamefont {Enss}},\ and\ \bibinfo {author}
  {\bibfnamefont {N.}~\bibnamefont {Defenu}},\ }\bibfield  {title} {\bibinfo
  {title} {Complex networks with tuneable spectral dimension as a universality
  playground},\ }\href {https://doi.org/10.1103/PhysRevResearch.3.023015}
  {\bibfield  {journal} {\bibinfo  {journal} {Phys. Rev. Research}\ }\textbf
  {\bibinfo {volume} {3}},\ \bibinfo {pages} {023015} (\bibinfo {year}
  {2021})}\BibitemShut {NoStop}%
\bibitem [{\citenamefont {Sarkar}\ \emph {et~al.}(2024)\citenamefont {Sarkar},
  \citenamefont {Enss},\ and\ \citenamefont {Defenu}}]{sarkar2024universality}%
  \BibitemOpen
  \bibfield  {author} {\bibinfo {author} {\bibfnamefont {M.}~\bibnamefont
  {Sarkar}}, \bibinfo {author} {\bibfnamefont {T.}~\bibnamefont {Enss}},\ and\
  \bibinfo {author} {\bibfnamefont {N.}~\bibnamefont {Defenu}},\ }\bibfield
  {title} {\bibinfo {title} {Universality of critical dynamics on a complex
  network},\ }\href {https://doi.org/10.1103/PhysRevB.110.014208} {\bibfield
  {journal} {\bibinfo  {journal} {Phys. Rev. B}\ }\textbf {\bibinfo {volume}
  {110}},\ \bibinfo {pages} {014208} (\bibinfo {year} {2024})}\BibitemShut
  {NoStop}%
\bibitem [{\citenamefont {Vojta}(2019)}]{vojta2019disorder}%
  \BibitemOpen
  \bibfield  {author} {\bibinfo {author} {\bibfnamefont {T.}~\bibnamefont
  {Vojta}},\ }\bibfield  {title} {\bibinfo {title} {Disorder in quantum
  many-body systems},\ }\href
  {https://doi.org/10.1146/annurev-conmatphys-031218-013433} {\bibfield
  {journal} {\bibinfo  {journal} {Annual Review of Condensed Matter Physics}\
  }\textbf {\bibinfo {volume} {10}},\ \bibinfo {pages} {233} (\bibinfo {year}
  {2019})}\BibitemShut {NoStop}%
\bibitem [{\citenamefont {Dahlberg}\ \emph {et~al.}(2025)\citenamefont
  {Dahlberg}, \citenamefont {Gonz{\'a}lez-Adalid~Pemart{\'\i}n}, \citenamefont
  {Marinari}, \citenamefont {Parisi}, \citenamefont {Ricci-Tersenghi},
  \citenamefont {Martin-Mayor}, \citenamefont {Moreno-Gordo}, \citenamefont
  {Orbach}, \citenamefont {Paga}, \citenamefont {Ruiz-Lorenzo} \emph
  {et~al.}}]{dahlberg2025spin}%
  \BibitemOpen
  \bibfield  {author} {\bibinfo {author} {\bibfnamefont {E.}~\bibnamefont
  {Dahlberg}}, \bibinfo {author} {\bibfnamefont {I.}~\bibnamefont
  {Gonz{\'a}lez-Adalid~Pemart{\'\i}n}}, \bibinfo {author} {\bibfnamefont
  {E.}~\bibnamefont {Marinari}}, \bibinfo {author} {\bibfnamefont
  {G.}~\bibnamefont {Parisi}}, \bibinfo {author} {\bibfnamefont
  {F.}~\bibnamefont {Ricci-Tersenghi}}, \bibinfo {author} {\bibfnamefont
  {V.}~\bibnamefont {Martin-Mayor}}, \bibinfo {author} {\bibfnamefont
  {J.}~\bibnamefont {Moreno-Gordo}}, \bibinfo {author} {\bibfnamefont
  {R.}~\bibnamefont {Orbach}}, \bibinfo {author} {\bibfnamefont
  {I.}~\bibnamefont {Paga}}, \bibinfo {author} {\bibfnamefont {J.}~\bibnamefont
  {Ruiz-Lorenzo}}, \emph {et~al.},\ }\bibfield  {title} {\bibinfo {title}
  {Spin-glass dynamics: Experiment, theory, and simulation},\ }\href
  {https://doi.org/https://doi.org/10.1103/ctp2-zwyr} {\bibfield  {journal}
  {\bibinfo  {journal} {Rev. Mod. Phys.}\ }\textbf {\bibinfo {volume} {97}},\
  \bibinfo {pages} {045005} (\bibinfo {year} {2025})}\BibitemShut {NoStop}%
\bibitem [{\citenamefont {Harris}(1974)}]{harris1974effect}%
  \BibitemOpen
  \bibfield  {author} {\bibinfo {author} {\bibfnamefont {A.~B.}\ \bibnamefont
  {Harris}},\ }\bibfield  {title} {\bibinfo {title} {Effect of random defects
  on the critical behaviour of ising models},\ }\href
  {https://doi.org/10.1088/0022-3719/7/9/009} {\bibfield  {journal} {\bibinfo
  {journal} {J. Phys. C: Solid State Phys.}\ }\textbf {\bibinfo {volume} {7}},\
  \bibinfo {pages} {1671} (\bibinfo {year} {1974})}\BibitemShut {NoStop}%
\bibitem [{\citenamefont {Weinrib}\ and\ \citenamefont
  {Halperin}(1983)}]{weinrib1983critical}%
  \BibitemOpen
  \bibfield  {author} {\bibinfo {author} {\bibfnamefont {A.}~\bibnamefont
  {Weinrib}}\ and\ \bibinfo {author} {\bibfnamefont {B.~I.}\ \bibnamefont
  {Halperin}},\ }\bibfield  {title} {\bibinfo {title} {Critical phenomena in
  systems with long-range-correlated quenched disorder},\ }\href
  {https://doi.org/https://doi.org/10.1103/PhysRevB.27.413} {\bibfield
  {journal} {\bibinfo  {journal} {Phys. Rev. B}\ }\textbf {\bibinfo {volume}
  {27}},\ \bibinfo {pages} {413} (\bibinfo {year} {1983})}\BibitemShut
  {NoStop}%
\bibitem [{\citenamefont {Liu}\ \emph {et~al.}(1999)\citenamefont {Liu},
  \citenamefont {Chen},\ and\ \citenamefont {Xiong}}]{liu1999kosterlitz}%
  \BibitemOpen
  \bibfield  {author} {\bibinfo {author} {\bibfnamefont {W.-S.}\ \bibnamefont
  {Liu}}, \bibinfo {author} {\bibfnamefont {T.}~\bibnamefont {Chen}},\ and\
  \bibinfo {author} {\bibfnamefont {S.-J.}\ \bibnamefont {Xiong}},\ }\bibfield
  {title} {\bibinfo {title} {Kosterlitz-thouless-type metal-insulator
  transition in two-dimensional layered media with long-range correlated
  disorder},\ }\href {https://doi.org/10.1088/0953-8984/11/36/306} {\bibfield
  {journal} {\bibinfo  {journal} {J. Phys.: Condens. Matter}\ }\textbf
  {\bibinfo {volume} {11}},\ \bibinfo {pages} {6883} (\bibinfo {year}
  {1999})}\BibitemShut {NoStop}%
\bibitem [{\citenamefont {Rieger}\ and\ \citenamefont
  {Igl{\'o}i}(1999)}]{rieger1999random}%
  \BibitemOpen
  \bibfield  {author} {\bibinfo {author} {\bibfnamefont {H.}~\bibnamefont
  {Rieger}}\ and\ \bibinfo {author} {\bibfnamefont {F.}~\bibnamefont
  {Igl{\'o}i}},\ }\bibfield  {title} {\bibinfo {title} {Random quantum magnets
  with long-range correlated disorder: Enhancement of critical and
  griffiths-mccoy singularities},\ }\href
  {https://doi.org/https://doi.org/10.1103/PhysRevLett.83.3741} {\bibfield
  {journal} {\bibinfo  {journal} {Phys. Rev. Lett.}\ }\textbf {\bibinfo
  {volume} {83}},\ \bibinfo {pages} {3741} (\bibinfo {year}
  {1999})}\BibitemShut {NoStop}%
\bibitem [{\citenamefont {Schrenk}\ \emph {et~al.}(2013)\citenamefont
  {Schrenk}, \citenamefont {Pos{\'e}}, \citenamefont {Kranz}, \citenamefont
  {Van~Kessenich}, \citenamefont {Ara{\'u}jo},\ and\ \citenamefont
  {Herrmann}}]{schrenk2013percolation}%
  \BibitemOpen
  \bibfield  {author} {\bibinfo {author} {\bibfnamefont {K.~J.}\ \bibnamefont
  {Schrenk}}, \bibinfo {author} {\bibfnamefont {N.}~\bibnamefont {Pos{\'e}}},
  \bibinfo {author} {\bibfnamefont {J.~J.}\ \bibnamefont {Kranz}}, \bibinfo
  {author} {\bibfnamefont {L.}~\bibnamefont {Van~Kessenich}}, \bibinfo {author}
  {\bibfnamefont {N.~A.}\ \bibnamefont {Ara{\'u}jo}},\ and\ \bibinfo {author}
  {\bibfnamefont {H.~J.}\ \bibnamefont {Herrmann}},\ }\bibfield  {title}
  {\bibinfo {title} {Percolation with long-range correlated disorder},\ }\href
  {https://doi.org/https://doi.org/10.1103/PhysRevE.88.052102} {\bibfield
  {journal} {\bibinfo  {journal} {Phys. Rev. E}\ }\textbf {\bibinfo {volume}
  {88}},\ \bibinfo {pages} {052102} (\bibinfo {year} {2013})}\BibitemShut
  {NoStop}%
\bibitem [{\citenamefont {Ibrahim}\ \emph {et~al.}(2014)\citenamefont
  {Ibrahim}, \citenamefont {Barghathi},\ and\ \citenamefont
  {Vojta}}]{ibrahim2014enhanced}%
  \BibitemOpen
  \bibfield  {author} {\bibinfo {author} {\bibfnamefont {A.~K.}\ \bibnamefont
  {Ibrahim}}, \bibinfo {author} {\bibfnamefont {H.}~\bibnamefont {Barghathi}},\
  and\ \bibinfo {author} {\bibfnamefont {T.}~\bibnamefont {Vojta}},\ }\bibfield
   {title} {\bibinfo {title} {Enhanced rare-region effects in the contact
  process with long-range correlated disorder},\ }\href
  {https://doi.org/https://doi.org/10.1103/PhysRevE.90.042132} {\bibfield
  {journal} {\bibinfo  {journal} {Phys. Rev. E}\ }\textbf {\bibinfo {volume}
  {90}},\ \bibinfo {pages} {042132} (\bibinfo {year} {2014})}\BibitemShut
  {NoStop}%
\bibitem [{\citenamefont {Mendes}\ \emph {et~al.}(2019)\citenamefont {Mendes},
  \citenamefont {Almeida}, \citenamefont {Lyra},\ and\ \citenamefont
  {de~Moura}}]{mendes2019localization}%
  \BibitemOpen
  \bibfield  {author} {\bibinfo {author} {\bibfnamefont {C.}~\bibnamefont
  {Mendes}}, \bibinfo {author} {\bibfnamefont {G.}~\bibnamefont {Almeida}},
  \bibinfo {author} {\bibfnamefont {M.}~\bibnamefont {Lyra}},\ and\ \bibinfo
  {author} {\bibfnamefont {F.}~\bibnamefont {de~Moura}},\ }\bibfield  {title}
  {\bibinfo {title} {Localization-delocalization transition in discrete-time
  quantum walks with long-range correlated disorder},\ }\href
  {https://doi.org/https://doi.org/10.1103/PhysRevE.99.022117} {\bibfield
  {journal} {\bibinfo  {journal} {Phys. Rev. E}\ }\textbf {\bibinfo {volume}
  {99}},\ \bibinfo {pages} {022117} (\bibinfo {year} {2019})}\BibitemShut
  {NoStop}%
\bibitem [{\citenamefont {Modak}\ and\ \citenamefont
  {Nag}(2020)}]{modak2020many}%
  \BibitemOpen
  \bibfield  {author} {\bibinfo {author} {\bibfnamefont {R.}~\bibnamefont
  {Modak}}\ and\ \bibinfo {author} {\bibfnamefont {T.}~\bibnamefont {Nag}},\
  }\bibfield  {title} {\bibinfo {title} {Many-body dynamics in long-range
  hopping models in the presence of correlated and uncorrelated disorder},\
  }\href {https://doi.org/https://doi.org/10.1103/PhysRevResearch.2.012074}
  {\bibfield  {journal} {\bibinfo  {journal} {Phys. Rev. Research}\ }\textbf
  {\bibinfo {volume} {2}},\ \bibinfo {pages} {012074} (\bibinfo {year}
  {2020})}\BibitemShut {NoStop}%
\bibitem [{\citenamefont {Ib{\'a}{\~n}ez~Berganza}\ and\ \citenamefont
  {Leuzzi}(2013)}]{berganza2013critical}%
  \BibitemOpen
  \bibfield  {author} {\bibinfo {author} {\bibfnamefont {M.}~\bibnamefont
  {Ib{\'a}{\~n}ez~Berganza}}\ and\ \bibinfo {author} {\bibfnamefont
  {L.}~\bibnamefont {Leuzzi}},\ }\bibfield  {title} {\bibinfo {title}
  {{Critical behavior of the XY model in complex topologies}},\ }\href
  {https://doi.org/https://doi.org/10.1103/PhysRevB.88.144104} {\bibfield
  {journal} {\bibinfo  {journal} {Phys. Rev. B}\ }\textbf {\bibinfo {volume}
  {88}},\ \bibinfo {pages} {144104} (\bibinfo {year} {2013})}\BibitemShut
  {NoStop}%
\bibitem [{\citenamefont {Cescatti}\ \emph {et~al.}(2019)\citenamefont
  {Cescatti}, \citenamefont {Ib{\'a}{\~n}ez-Berganza}, \citenamefont
  {Vezzani},\ and\ \citenamefont {Burioni}}]{cescatti2019analysis}%
  \BibitemOpen
  \bibfield  {author} {\bibinfo {author} {\bibfnamefont {F.}~\bibnamefont
  {Cescatti}}, \bibinfo {author} {\bibfnamefont {M.}~\bibnamefont
  {Ib{\'a}{\~n}ez-Berganza}}, \bibinfo {author} {\bibfnamefont
  {A.}~\bibnamefont {Vezzani}},\ and\ \bibinfo {author} {\bibfnamefont
  {R.}~\bibnamefont {Burioni}},\ }\bibfield  {title} {\bibinfo {title}
  {{Analysis of the low-temperature phase in the two-dimensional long-range
  diluted XY model}},\ }\href
  {https://doi.org/https://doi.org/10.1103/PhysRevB.100.054203} {\bibfield
  {journal} {\bibinfo  {journal} {Phys. Rev. B}\ }\textbf {\bibinfo {volume}
  {100}},\ \bibinfo {pages} {054203} (\bibinfo {year} {2019})}\BibitemShut
  {NoStop}%
\bibitem [{\citenamefont {Bighin}\ \emph {et~al.}(2024)\citenamefont {Bighin},
  \citenamefont {Enss},\ and\ \citenamefont {Defenu}}]{bighin2024universal}%
  \BibitemOpen
  \bibfield  {author} {\bibinfo {author} {\bibfnamefont {G.}~\bibnamefont
  {Bighin}}, \bibinfo {author} {\bibfnamefont {T.}~\bibnamefont {Enss}},\ and\
  \bibinfo {author} {\bibfnamefont {N.}~\bibnamefont {Defenu}},\ }\bibfield
  {title} {\bibinfo {title} {Universal scaling in real dimension},\ }\href
  {https://doi.org/https://doi.org/10.1038/s41467-024-48537-1} {\bibfield
  {journal} {\bibinfo  {journal} {Nat. Commun.}\ }\textbf {\bibinfo {volume}
  {15}},\ \bibinfo {pages} {4207} (\bibinfo {year} {2024})}\BibitemShut
  {NoStop}%
\bibitem [{\citenamefont {Leuzzi}\ \emph {et~al.}(2008)\citenamefont {Leuzzi},
  \citenamefont {Parisi}, \citenamefont {Ricci-Tersenghi},\ and\ \citenamefont
  {Ruiz-Lorenzo}}]{leuzzi2008dilute}%
  \BibitemOpen
  \bibfield  {author} {\bibinfo {author} {\bibfnamefont {L.}~\bibnamefont
  {Leuzzi}}, \bibinfo {author} {\bibfnamefont {G.}~\bibnamefont {Parisi}},
  \bibinfo {author} {\bibfnamefont {F.}~\bibnamefont {Ricci-Tersenghi}},\ and\
  \bibinfo {author} {\bibfnamefont {J.~J.}\ \bibnamefont {Ruiz-Lorenzo}},\
  }\bibfield  {title} {\bibinfo {title} {Dilute one-dimensional spin glasses
  with power law decaying interactions},\ }\href
  {https://doi.org/10.1103/PhysRevLett.101.107203} {\bibfield  {journal}
  {\bibinfo  {journal} {Phys. Rev. Lett.}\ }\textbf {\bibinfo {volume} {101}},\
  \bibinfo {pages} {107203} (\bibinfo {year} {2008})}\BibitemShut {NoStop}%
\bibitem [{\citenamefont {Sarkar}\ \emph {et~al.}(2025)\citenamefont {Sarkar},
  \citenamefont {Defenu},\ and\ \citenamefont {Enss}}]{sarkar2025long}%
  \BibitemOpen
  \bibfield  {author} {\bibinfo {author} {\bibfnamefont {M.}~\bibnamefont
  {Sarkar}}, \bibinfo {author} {\bibfnamefont {N.}~\bibnamefont {Defenu}},\
  and\ \bibinfo {author} {\bibfnamefont {T.}~\bibnamefont {Enss}},\ }\bibfield
  {title} {\bibinfo {title} {Long range to short range crossover in one
  dimension},\ }\href {https://doi.org/10.48550/arXiv.2507.08092} {\bibfield
  {journal} {\bibinfo  {journal} {arXiv:2507.08092}\ } (\bibinfo {year}
  {2025})}\BibitemShut {NoStop}%
\bibitem [{\citenamefont {Grassberger}(1985)}]{grassberger1985critical}%
  \BibitemOpen
  \bibfield  {author} {\bibinfo {author} {\bibfnamefont {P.}~\bibnamefont
  {Grassberger}},\ }\bibfield  {title} {\bibinfo {title} {Critical exponents of
  self-avoiding {L{\'e}vy} flights},\ }\href
  {https://doi.org/10.1088/0305-4470/18/8/011} {\bibfield  {journal} {\bibinfo
  {journal} {J. Phys. A: Math. Gen.}\ }\textbf {\bibinfo {volume} {18}},\
  \bibinfo {pages} {L463} (\bibinfo {year} {1985})}\BibitemShut {NoStop}%
\bibitem [{\citenamefont {Sokal}(1994)}]{sokal1994monte}%
  \BibitemOpen
  \bibfield  {author} {\bibinfo {author} {\bibfnamefont {A.~D.}\ \bibnamefont
  {Sokal}},\ }\bibfield  {title} {\bibinfo {title} {Monte carlo methods for the
  self-avoiding walk},\ }\href {https://doi.org/10.48550/arXiv.hep-lat/9509032}
  {\bibfield  {journal} {\bibinfo  {journal} {arXiv preprint hep-lat/9405016}\
  } (\bibinfo {year} {1994})}\BibitemShut {NoStop}%
\bibitem [{\citenamefont {de~Gennes}(1972)}]{de1972exponents}%
  \BibitemOpen
  \bibfield  {author} {\bibinfo {author} {\bibfnamefont {P.-G.}\ \bibnamefont
  {de~Gennes}},\ }\bibfield  {title} {\bibinfo {title} {Exponents for the
  excluded volume problem as derived by the wilson method},\ }\href
  {https://doi.org/10.1016/0375-9601(72)90149-1} {\bibfield  {journal}
  {\bibinfo  {journal} {Phys. Lett. A}\ }\textbf {\bibinfo {volume} {38}},\
  \bibinfo {pages} {339} (\bibinfo {year} {1972})}\BibitemShut {NoStop}%
\bibitem [{\citenamefont {Nishimori}\ and\ \citenamefont
  {Ortiz}(2010)}]{nishimori2010elements}%
  \BibitemOpen
  \bibfield  {author} {\bibinfo {author} {\bibfnamefont {H.}~\bibnamefont
  {Nishimori}}\ and\ \bibinfo {author} {\bibfnamefont {G.}~\bibnamefont
  {Ortiz}},\ }\href@noop {} {\emph {\bibinfo {title} {Elements of phase
  transitions and critical phenomena}}}\ (\bibinfo  {publisher} {Oup Oxford},\
  \bibinfo {year} {2010})\BibitemShut {NoStop}%
\bibitem [{\citenamefont {Machado}\ and\ \citenamefont
  {Dupuis}(2010)}]{machado2010local}%
  \BibitemOpen
  \bibfield  {author} {\bibinfo {author} {\bibfnamefont {T.}~\bibnamefont
  {Machado}}\ and\ \bibinfo {author} {\bibfnamefont {N.}~\bibnamefont
  {Dupuis}},\ }\bibfield  {title} {\bibinfo {title} {From local to critical
  fluctuations in lattice models: A nonperturbative renormalization-group
  approach},\ }\href
  {https://doi.org/https://doi.org/10.1103/PhysRevE.82.041128} {\bibfield
  {journal} {\bibinfo  {journal} {Phys. Rev. E}\ }\textbf {\bibinfo {volume}
  {82}},\ \bibinfo {pages} {041128} (\bibinfo {year} {2010})}\BibitemShut
  {NoStop}%
\bibitem [{SM()}]{SM}%
  \BibitemOpen
  \href@noop {} {}\bibinfo {note} {See Supplemental Material at THIS URL for
  details on the full derivation.}\BibitemShut {Stop}%
\bibitem [{\citenamefont {Dotsenko}(2005)}]{dotsenko2005introduction}%
  \BibitemOpen
  \bibfield  {author} {\bibinfo {author} {\bibfnamefont {V.}~\bibnamefont
  {Dotsenko}},\ }\bibfield  {title} {\bibinfo {title} {Introduction to the
  replica theory of disordered statistical systems},\ }\href@noop {} {\bibfield
   {journal} {\bibinfo  {journal} {Introduction to the Replica Theory of
  Disordered Statistical Systems}\ } (\bibinfo {year} {2005})}\BibitemShut
  {NoStop}%
\bibitem [{\citenamefont {Chakrabarti}\ and\ \citenamefont
  {Kertesz}(1981)}]{chakrabarti1981statistics}%
  \BibitemOpen
  \bibfield  {author} {\bibinfo {author} {\bibfnamefont {B.}~\bibnamefont
  {Chakrabarti}}\ and\ \bibinfo {author} {\bibfnamefont {J.}~\bibnamefont
  {Kertesz}},\ }\bibfield  {title} {\bibinfo {title} {The statistics of
  self-avoiding walks on a disordered lattice},\ }\href
  {https://doi.org/https://doi.org/10.1007/BF01297178} {\bibfield  {journal}
  {\bibinfo  {journal} {Zeitschrift f{\"u}r Physik B Condensed Matter}\
  }\textbf {\bibinfo {volume} {44}},\ \bibinfo {pages} {221} (\bibinfo {year}
  {1981})}\BibitemShut {NoStop}%
\bibitem [{\citenamefont {Kim}(1983)}]{kim1983renormalisation}%
  \BibitemOpen
  \bibfield  {author} {\bibinfo {author} {\bibfnamefont {Y.}~\bibnamefont
  {Kim}},\ }\bibfield  {title} {\bibinfo {title} {Renormalisation-group study
  of self-avoiding walks on the random lattice},\ }\href
  {https://doi.org/10.1088/0022-3719/16/8/005} {\bibfield  {journal} {\bibinfo
  {journal} {J. Phys. C: Solid State Phys.}\ }\textbf {\bibinfo {volume}
  {16}},\ \bibinfo {pages} {1345} (\bibinfo {year} {1983})}\BibitemShut
  {NoStop}%
\bibitem [{\citenamefont {Meir}\ and\ \citenamefont
  {Harris}(1989)}]{meir1989self}%
  \BibitemOpen
  \bibfield  {author} {\bibinfo {author} {\bibfnamefont {Y.}~\bibnamefont
  {Meir}}\ and\ \bibinfo {author} {\bibfnamefont {A.~B.}\ \bibnamefont
  {Harris}},\ }\bibfield  {title} {\bibinfo {title} {Self-avoiding walks on
  diluted networks},\ }\href
  {https://doi.org/https://doi.org/10.1103/PhysRevLett.63.2819} {\bibfield
  {journal} {\bibinfo  {journal} {Phys. Rev. Lett.}\ }\textbf {\bibinfo
  {volume} {63}},\ \bibinfo {pages} {2819} (\bibinfo {year}
  {1989})}\BibitemShut {NoStop}%
\bibitem [{\citenamefont {Grassberger}(1993)}]{grassberger1993recursive}%
  \BibitemOpen
  \bibfield  {author} {\bibinfo {author} {\bibfnamefont {P.}~\bibnamefont
  {Grassberger}},\ }\bibfield  {title} {\bibinfo {title} {Recursive sampling of
  random walks: self-avoiding walks in disordered media},\ }\href
  {https://doi.org/10.1088/0305-4470/26/5/022} {\bibfield  {journal} {\bibinfo
  {journal} {J. Phys. A: Math. Gen.}\ }\textbf {\bibinfo {volume} {26}},\
  \bibinfo {pages} {1023} (\bibinfo {year} {1993})}\BibitemShut {NoStop}%
\bibitem [{\citenamefont {Barat}\ and\ \citenamefont
  {Chakrabarti}(1995)}]{barat1995statistics}%
  \BibitemOpen
  \bibfield  {author} {\bibinfo {author} {\bibfnamefont {K.}~\bibnamefont
  {Barat}}\ and\ \bibinfo {author} {\bibfnamefont {B.~K.}\ \bibnamefont
  {Chakrabarti}},\ }\bibfield  {title} {\bibinfo {title} {Statistics of
  self-avoiding walks on random lattices},\ }\href
  {https://doi.org/https://doi.org/10.1016/0370-1573(95)00009-6} {\bibfield
  {journal} {\bibinfo  {journal} {Physics Reports}\ }\textbf {\bibinfo {volume}
  {258}},\ \bibinfo {pages} {377} (\bibinfo {year} {1995})}\BibitemShut
  {NoStop}%
\bibitem [{\citenamefont {Kim}(1987)}]{kim1987self}%
  \BibitemOpen
  \bibfield  {author} {\bibinfo {author} {\bibfnamefont {Y.}~\bibnamefont
  {Kim}},\ }\bibfield  {title} {\bibinfo {title} {Self-avoiding walks on
  lattices with a long-range-correlated disorder},\ }\href
  {https://doi.org/10.1088/0305-4470/20/17/037} {\bibfield  {journal} {\bibinfo
   {journal} {J. Phys. A: Math. Gen.}\ }\textbf {\bibinfo {volume} {20}},\
  \bibinfo {pages} {6047} (\bibinfo {year} {1987})}\BibitemShut {NoStop}%
\bibitem [{\citenamefont {Cardy}(1996)}]{cardy1996scaling}%
  \BibitemOpen
  \bibfield  {author} {\bibinfo {author} {\bibfnamefont {J.}~\bibnamefont
  {Cardy}},\ }\href@noop {} {\emph {\bibinfo {title} {Scaling and
  renormalization in statistical physics}}},\ Vol.~\bibinfo {volume} {5}\
  (\bibinfo  {publisher} {Cambridge university press},\ \bibinfo {year}
  {1996})\BibitemShut {NoStop}%
\bibitem [{\citenamefont {Sak}(1973)}]{sak1973recursion}%
  \BibitemOpen
  \bibfield  {author} {\bibinfo {author} {\bibfnamefont {J.}~\bibnamefont
  {Sak}},\ }\bibfield  {title} {\bibinfo {title} {Recursion relations and fixed
  points for ferromagnets with long-range interactions},\ }\href
  {https://doi.org/https://doi.org/10.1103/PhysRevB.8.281} {\bibfield
  {journal} {\bibinfo  {journal} {Phys. Rev. B}\ }\textbf {\bibinfo {volume}
  {8}},\ \bibinfo {pages} {281} (\bibinfo {year} {1973})}\BibitemShut {NoStop}%
\bibitem [{\citenamefont {Brezin}\ \emph {et~al.}(2014)\citenamefont {Brezin},
  \citenamefont {Parisi},\ and\ \citenamefont
  {Ricci-Tersenghi}}]{brezin2014crossover}%
  \BibitemOpen
  \bibfield  {author} {\bibinfo {author} {\bibfnamefont {E.}~\bibnamefont
  {Brezin}}, \bibinfo {author} {\bibfnamefont {G.}~\bibnamefont {Parisi}},\
  and\ \bibinfo {author} {\bibfnamefont {F.}~\bibnamefont {Ricci-Tersenghi}},\
  }\bibfield  {title} {\bibinfo {title} {The crossover region between
  long-range and short-range interactions for the critical exponents},\ }\href
  {https://doi.org/https://doi.org/10.1007/s10955-014-1081-0} {\bibfield
  {journal} {\bibinfo  {journal} {J. Stat. Phys.}\ }\textbf {\bibinfo {volume}
  {157}},\ \bibinfo {pages} {855} (\bibinfo {year} {2014})}\BibitemShut
  {NoStop}%
\bibitem [{\citenamefont {Defenu}\ \emph {et~al.}(2015)\citenamefont {Defenu},
  \citenamefont {Trombettoni},\ and\ \citenamefont
  {Codello}}]{defenu2015fixed}%
  \BibitemOpen
  \bibfield  {author} {\bibinfo {author} {\bibfnamefont {N.}~\bibnamefont
  {Defenu}}, \bibinfo {author} {\bibfnamefont {A.}~\bibnamefont
  {Trombettoni}},\ and\ \bibinfo {author} {\bibfnamefont {A.}~\bibnamefont
  {Codello}},\ }\bibfield  {title} {\bibinfo {title} {Fixed-point structure and
  effective fractional dimensionality for {$O (N)$} models with long-range
  interactions},\ }\href
  {https://doi.org/https://doi.org/10.1103/PhysRevE.92.052113} {\bibfield
  {journal} {\bibinfo  {journal} {Phys. Rev. E}\ }\textbf {\bibinfo {volume}
  {92}},\ \bibinfo {pages} {052113} (\bibinfo {year} {2015})}\BibitemShut
  {NoStop}%
\bibitem [{\citenamefont {Behan}\ \emph
  {et~al.}(2017{\natexlab{a}})\citenamefont {Behan}, \citenamefont {Rastelli},
  \citenamefont {Rychkov},\ and\ \citenamefont {Zan}}]{behan2017long}%
  \BibitemOpen
  \bibfield  {author} {\bibinfo {author} {\bibfnamefont {C.}~\bibnamefont
  {Behan}}, \bibinfo {author} {\bibfnamefont {L.}~\bibnamefont {Rastelli}},
  \bibinfo {author} {\bibfnamefont {S.}~\bibnamefont {Rychkov}},\ and\ \bibinfo
  {author} {\bibfnamefont {B.}~\bibnamefont {Zan}},\ }\bibfield  {title}
  {\bibinfo {title} {Long-range critical exponents near the short-range
  crossover},\ }\href
  {https://doi.org/https://doi.org/10.1103/PhysRevLett.118.241601 Export
  Citation} {\bibfield  {journal} {\bibinfo  {journal} {Phys. Rev. Lett.}\
  }\textbf {\bibinfo {volume} {118}},\ \bibinfo {pages} {241601} (\bibinfo
  {year} {2017}{\natexlab{a}})}\BibitemShut {NoStop}%
\bibitem [{\citenamefont {Behan}\ \emph
  {et~al.}(2017{\natexlab{b}})\citenamefont {Behan}, \citenamefont {Rastelli},
  \citenamefont {Rychkov},\ and\ \citenamefont {Zan}}]{behan2017scaling}%
  \BibitemOpen
  \bibfield  {author} {\bibinfo {author} {\bibfnamefont {C.}~\bibnamefont
  {Behan}}, \bibinfo {author} {\bibfnamefont {L.}~\bibnamefont {Rastelli}},
  \bibinfo {author} {\bibfnamefont {S.}~\bibnamefont {Rychkov}},\ and\ \bibinfo
  {author} {\bibfnamefont {B.}~\bibnamefont {Zan}},\ }\bibfield  {title}
  {\bibinfo {title} {A scaling theory for the long-range to short-range
  crossover and an infrared duality},\ }\href
  {https://doi.org/10.1088/1751-8121/aa8099} {\bibfield  {journal} {\bibinfo
  {journal} {J. Phys. A: Math. Theor.}\ }\textbf {\bibinfo {volume} {50}},\
  \bibinfo {pages} {354002} (\bibinfo {year} {2017}{\natexlab{b}})}\BibitemShut
  {NoStop}%
\bibitem [{\citenamefont {Benedetti}\ \emph {et~al.}(2025)\citenamefont
  {Benedetti}, \citenamefont {Lauria}, \citenamefont {Maz{\'a}{\v{c}}},\ and\
  \citenamefont {van Vliet}}]{benedetti2025one}%
  \BibitemOpen
  \bibfield  {author} {\bibinfo {author} {\bibfnamefont {D.}~\bibnamefont
  {Benedetti}}, \bibinfo {author} {\bibfnamefont {E.}~\bibnamefont {Lauria}},
  \bibinfo {author} {\bibfnamefont {D.}~\bibnamefont {Maz{\'a}{\v{c}}}},\ and\
  \bibinfo {author} {\bibfnamefont {P.}~\bibnamefont {van Vliet}},\ }\bibfield
  {title} {\bibinfo {title} {One-dimensional ising model with $1/r^{1.99}$
  interaction},\ }\href
  {https://doi.org/https://doi.org/10.1103/PhysRevLett.134.201602} {\bibfield
  {journal} {\bibinfo  {journal} {Phys. Rev. Lett.}\ }\textbf {\bibinfo
  {volume} {134}},\ \bibinfo {pages} {201602} (\bibinfo {year}
  {2025})}\BibitemShut {NoStop}%
\bibitem [{\citenamefont {Li}\ \emph {et~al.}(2026)\citenamefont {Li},
  \citenamefont {Chen},\ and\ \citenamefont {Deng}}]{li20264epsilon}%
  \BibitemOpen
  \bibfield  {author} {\bibinfo {author} {\bibfnamefont {Z.}~\bibnamefont
  {Li}}, \bibinfo {author} {\bibfnamefont {K.}~\bibnamefont {Chen}},\ and\
  \bibinfo {author} {\bibfnamefont {Y.}~\bibnamefont {Deng}},\ }\href
  {https://arxiv.org/abs/2602.07818} {\bibinfo {title} {The 4-$\epsilon$
  expansion for long-range interacting systems}} (\bibinfo {year} {2026}),\
  \Eprint {https://arxiv.org/abs/2602.07818} {arXiv:2602.07818} \BibitemShut
  {NoStop}%
\bibitem [{\citenamefont {Halley}\ and\ \citenamefont
  {Nakanishi}(1985)}]{halley1985node}%
  \BibitemOpen
  \bibfield  {author} {\bibinfo {author} {\bibfnamefont {J.}~\bibnamefont
  {Halley}}\ and\ \bibinfo {author} {\bibfnamefont {H.}~\bibnamefont
  {Nakanishi}},\ }\bibfield  {title} {\bibinfo {title} {Node-avoiding l{\'e}vy
  flight: a numerical test of the $\varepsilon$ expansion},\ }\href
  {https://doi.org/https://doi.org/10.1103/PhysRevLett.55.551} {\bibfield
  {journal} {\bibinfo  {journal} {Phys. Rev. Lett.}\ }\textbf {\bibinfo
  {volume} {55}},\ \bibinfo {pages} {551} (\bibinfo {year} {1985})}\BibitemShut
  {NoStop}%
\bibitem [{\citenamefont {De~Gennes}(1979)}]{de1979scaling}%
  \BibitemOpen
  \bibfield  {author} {\bibinfo {author} {\bibfnamefont {P.-G.}\ \bibnamefont
  {De~Gennes}},\ }\href@noop {} {\emph {\bibinfo {title} {Scaling concepts in
  polymer physics}}}\ (\bibinfo  {publisher} {Cornell University Press},\
  \bibinfo {year} {1979})\BibitemShut {NoStop}%
\bibitem [{\citenamefont {Sarkar}\ \emph {et~al.}(2026)\citenamefont {Sarkar},
  \citenamefont {Pagni}, \citenamefont {Enss},\ and\ \citenamefont
  {Defenu}}]{sarkar2026emergent}%
  \BibitemOpen
  \bibfield  {author} {\bibinfo {author} {\bibfnamefont {M.}~\bibnamefont
  {Sarkar}}, \bibinfo {author} {\bibfnamefont {V.}~\bibnamefont {Pagni}},
  \bibinfo {author} {\bibfnamefont {T.}~\bibnamefont {Enss}},\ and\ \bibinfo
  {author} {\bibfnamefont {N.}~\bibnamefont {Defenu}},\ }\bibfield  {title}
  {\bibinfo {title} {Emergent quantum chaos from correlations on a random
  graph},\ }\href {https://doi.org/10.48550/arXiv.2607.11662} {\bibfield
  {journal} {\bibinfo  {journal} {arXiv:2607.11662}\ } (\bibinfo {year}
  {2026})}\BibitemShut {NoStop}%
\bibitem [{\citenamefont {Qiang}\ \emph {et~al.}(2021)\citenamefont {Qiang},
  \citenamefont {Wang}, \citenamefont {Xue}, \citenamefont {Ge}, \citenamefont
  {Chen}, \citenamefont {Liu}, \citenamefont {Huang}, \citenamefont {Fu},
  \citenamefont {Xu}, \citenamefont {Yi}, \citenamefont {Xu}, \citenamefont
  {Deng}, \citenamefont {Wang}, \citenamefont {Meinecke}, \citenamefont
  {Matthews}, \citenamefont {Cai}, \citenamefont {Yang},\ and\ \citenamefont
  {Wu}}]{qiang2021implementing}%
  \BibitemOpen
  \bibfield  {author} {\bibinfo {author} {\bibfnamefont {X.}~\bibnamefont
  {Qiang}}, \bibinfo {author} {\bibfnamefont {Y.}~\bibnamefont {Wang}},
  \bibinfo {author} {\bibfnamefont {S.}~\bibnamefont {Xue}}, \bibinfo {author}
  {\bibfnamefont {R.}~\bibnamefont {Ge}}, \bibinfo {author} {\bibfnamefont
  {L.}~\bibnamefont {Chen}}, \bibinfo {author} {\bibfnamefont {Y.}~\bibnamefont
  {Liu}}, \bibinfo {author} {\bibfnamefont {A.}~\bibnamefont {Huang}}, \bibinfo
  {author} {\bibfnamefont {X.}~\bibnamefont {Fu}}, \bibinfo {author}
  {\bibfnamefont {P.}~\bibnamefont {Xu}}, \bibinfo {author} {\bibfnamefont
  {T.}~\bibnamefont {Yi}}, \bibinfo {author} {\bibfnamefont {F.}~\bibnamefont
  {Xu}}, \bibinfo {author} {\bibfnamefont {M.}~\bibnamefont {Deng}}, \bibinfo
  {author} {\bibfnamefont {J.~B.}\ \bibnamefont {Wang}}, \bibinfo {author}
  {\bibfnamefont {J.~D.~A.}\ \bibnamefont {Meinecke}}, \bibinfo {author}
  {\bibfnamefont {J.~C.~F.}\ \bibnamefont {Matthews}}, \bibinfo {author}
  {\bibfnamefont {X.}~\bibnamefont {Cai}}, \bibinfo {author} {\bibfnamefont
  {X.}~\bibnamefont {Yang}},\ and\ \bibinfo {author} {\bibfnamefont
  {J.}~\bibnamefont {Wu}},\ }\bibfield  {title} {\bibinfo {title} {Implementing
  graph-theoretic quantum algorithms on a silicon photonic quantum walk
  processor},\ }\href {https://doi.org/10.1126/sciadv.abb8375} {\bibfield
  {journal} {\bibinfo  {journal} {Science Advances}\ }\textbf {\bibinfo
  {volume} {7}},\ \bibinfo {pages} {eabb8375} (\bibinfo {year}
  {2021})}\BibitemShut {NoStop}%
\bibitem [{\citenamefont {Nguyen}\ \emph {et~al.}(2023)\citenamefont {Nguyen},
  \citenamefont {Liu}, \citenamefont {Wurtz}, \citenamefont {Lukin},
  \citenamefont {Wang},\ and\ \citenamefont {Pichler}}]{nguyen2023quantum}%
  \BibitemOpen
  \bibfield  {author} {\bibinfo {author} {\bibfnamefont {M.-T.}\ \bibnamefont
  {Nguyen}}, \bibinfo {author} {\bibfnamefont {J.-G.}\ \bibnamefont {Liu}},
  \bibinfo {author} {\bibfnamefont {J.}~\bibnamefont {Wurtz}}, \bibinfo
  {author} {\bibfnamefont {M.~D.}\ \bibnamefont {Lukin}}, \bibinfo {author}
  {\bibfnamefont {S.-T.}\ \bibnamefont {Wang}},\ and\ \bibinfo {author}
  {\bibfnamefont {H.}~\bibnamefont {Pichler}},\ }\bibfield  {title} {\bibinfo
  {title} {Quantum optimization with arbitrary connectivity using rydberg atom
  arrays},\ }\href {https://doi.org/10.1103/PRXQuantum.4.010316} {\bibfield
  {journal} {\bibinfo  {journal} {PRX Quantum}\ }\textbf {\bibinfo {volume}
  {4}},\ \bibinfo {pages} {010316} (\bibinfo {year} {2023})}\BibitemShut
  {NoStop}%
\end{thebibliography}%

\clearpage

\clearpage
\onecolumngrid  
\appendix
\begin{center}
	\textbf{\large Supplemental Material for ``Universality of superdiffusion in simple random graphs''}\\[5pt]
    Mrinal Sarkar$^1$, Nicolò Defenu$^2$, and Tilman Enss$^1$ \\[3pt]
    \textit{$^1$Institut für Theoretische Physik, Universität Heidelberg, 69120 Heidelberg, Germany} \\[3pt]
    \textit{$^2$Institut für Theoretische Physik, ETH Zürich, 8093 Zürich, Switzerland
} \\[3pt]
( Dated:\,\today)
\end{center}
\bigskip

\renewcommand{\theequation}{S\arabic{equation}}
\renewcommand{\thetable}{S\arabic{table}}
\renewcommand{\thefigure}{S\arabic{figure}}
\setcounter{equation}{0}
\setcounter{table}{0}
\setcounter{figure}{0}

\section{Mapping from discrete to continuous variable: Hubbard-Stratonovich transformation}
We consider the $O(n)$ model on the one dimensional long-range random ring graph (1DLR3). The Hamiltonian is defined by
\begin{equation}
    \beta H = - \frac{1}{2} \sum_{i \neq j} J_{ij} \vect{S}_i \cdot \vect{S}_j,  \qquad 
    \vect{S}_i \in \RR^n, \quad |\vect{S}_i|^2 = n,
\end{equation}
in the de Gennes formulation. The couplings $J_{ij}$ are independent random variables drawn from the following distribution
\begin{equation}
P(J_{ij}) = p_{ij} \, \delta(J_{ij} - 1) + (1 - p_{ij}) \, \delta(J_{ij}).
\label{eq:bond_dist_SI}
\end{equation}
Any link or bond between $(i,j)$ on the graph is connected with a probability $p_{ij}$ that decays as a power law,
\begin{equation}
p_{ij} = \frac{1}{|i-j|^{1+\sigma}},
\end{equation}
representing the long-range (LR) nature of the graph. The average degree is $\mathcal{O}(1)$, so $p_{ij} \ll 1$ for large $|i-j|$.

Define the $nN$-dimensional vector $\vect{S} = (\vect{S}_1, \vect{S}_2, \dots, \vect{S}_N)$, and an $nN \times nN $ symmetric coupling matrix $\vect{K'}$ to write the interaction term as a quadratic form:
\begin{align}
\frac{1}{2} \sum_{i \neq j} J_{ij} \vect{S}_i \cdot \vect{S}_j &= \frac{1}{2} \vect{S}^T \cdot \vect{K}' \cdot \vect{S}.
\end{align}
Since $\vect{K'}$ is not invertible, we introduce a shift $\vect{K} = \vect{K}' + 2 \mu \,\II_{nN}$, with the constant $\mu$ chosen to make the matrix positive definite. This shift corresponds to adding and subtracting a quadratic term in the spin, which can be absorbed into the potrential. Physically $\mu$ will contribute to the mass term. The partition function is
\begin{equation}
Z = \exp \left( - \mu n N \right)\int \left[\prod_i d\vect{S}_i \, \delta(|\vect{S}_i|^2 - n)\right] \exp\left(\frac{1}{2} \vect{S}^T \cdot \vect{K} \cdot \vect{S}\right).
\end{equation}
Now we apply Hubbard–Stratonovich transformation to map from discrete spin to a continuous field and decouple the the quadratic interaction. Introduce an auxiliary field $\vect{\phi}_i \in \RR^n$ and use the Gaussian identity,
\begin{equation}
\exp\left(\frac{1}{2} \vect{S}^T \cdot \vect{K} \cdot \vect{S}\right) = \frac{1}{\sqrt{\det(2\pi \vect{K}^{-1})}} \int \calD\vect{\phi} \, \exp\left(-\frac{1}{2} \vect{\phi}^T \cdot \vect{K}^{-1} \cdot \vect{\phi} - \vect{\phi}^T \cdot \vect{S}\right).
\end{equation}
The partition function becomes
\begin{equation}
Z =  \mathcal{N} \int \calD \vect{\phi}\, \exp\left(-\frac{1}{2} \vect{\phi}^T \cdot \vect{K}^{-1} \cdot \vect{\phi}  + \sum_j U (\vect{\phi}_j) \right)
\end{equation}
where $\mathcal{N}$ is a normalization constant, with the single-site potential
\begin{equation}
    e^{ U (\vect{\phi}_i)} =  \int d\vect{S}_j \, \delta(|\vect{S}_j|^2 - n) e^{- \vect{\phi}_i^T \cdot \vect{S}_j}
    \label{eq:U_phi}
\end{equation}
The integrand in Eq.\,\ref{eq:U_phi} depends only on $|\vect{\phi}_i|$. By rotational symmetry, we may take $\vect{\phi}_i$ along $\vect{S}_1$, and further define $\rho_i = |\vect{\phi}_i|^2$, yielding
\begin{equation}
e^{U(\vect{\rho})} = \int d \vect{S}_1 \, \delta(|\vect{S}_1|^2 - n) \exp(\sqrt{\rho} \vect{S}_1).
\end{equation}
Evaluating the integral finally yields
\begin{equation}
e^{U(\vect{\rho})} = f(n) \, (\sqrt{\rho})^{1- \frac{n}{2}} \, I_{\frac{n}{2} -1} (\sqrt{n \rho}).
\end{equation}
with $f(n) = n^{\frac{n}{4} - \frac{1}{2}} \, (2\pi)^{\frac{n}{2}}$ and $I_{\nu} (x)$ being the modified Bessel function of first kind.

\subsection{Quadratic term and random mass disorder}
The quadratic term in the partition function gives rise to derivatives in the field theory, and also contributes to the mass  term.
\begin{equation}
    \vect{\phi}^T \cdot  \vect{K}^{-1} \cdot \vect{\phi} = \sum_{i,j} \phi_i (K^{-1})_{ij} \phi_j = \sum_{i \neq j} \phi_i (K^{-1})_{ij} \phi_j + \sum_{i} (K^{-1})_{ii} \phi_i^2
\end{equation}
Now, by construction, $K_{ij} = J_{ij} + 2\mu \II \, \delta_{ij}$; also, by definition $J_{ij} = 0$ for $i=j$. In matrix notation,
\begin{equation}
K = 2\mu \, \II + J,
\end{equation}
Then,
\begin{equation}
K^{-1} = \frac{1}{2\mu} \left(\II + \frac{1}{2\mu} J\right)^{-1} = \frac{1}{2\mu} \sum_{m=0}^{\infty} \left(-\frac{J}{2\mu}\right)^m.
\end{equation}
The contribution to the effective mass term: (Keeping up to $O(J^2)$ for a sparse graph)
\begin{equation}
(K^{-1})_{ii} = \frac{1}{2\mu} \left[1 - \frac{1}{2\mu} J_{ii} + \frac{1}{4\mu^2} \sum_{k \neq i} J_{ik} J_{ik} + \dots \right] = \frac{1}{2\mu} \left[1 + \frac{1}{4\mu^2} \sum_{k \neq i} J_{ik}^2 \right] =  \frac{1}{2\mu} \left[1 + \frac{z_i}{4\mu^2} \right].
\end{equation}
Here we use the fact that $J_{ik}^2 = J_{ik}$ since $J_{ik} \in \{0,1\}$ and also identify the degree $z_i := \sum_{k \neq i} J_{ik}$ of the $i$-th node on the graph.

Thus the effective mass term at the $i$-th node on a given graph realization is
\begin{equation}
 = \frac{1}{2} \sum_i \left( -1 + \frac{1}{2\mu} + \frac{z_i}{8\mu^3} + \dots \right) |\phi_i|^2.
 \label{eq:supp_mass_terms}
\end{equation}
The degree $z_i$ in our graph is a random variable, we find that random bond disorder does generate random mass disorder.

\subsection{Emergent correlated mass disorder}
In this section we investigate the correlation of the mass disorder. By definition:
\begin{equation}
z_i = \sum_{j \neq i} J_{ij},\quad \bar{z} = \overline{z_i} = \sum_{j \neq i} \overline{J_{ij}} = \sum_{j \neq i} p_{ij},
\end{equation}
with $p_{ij} = |i-j|^{-1-\sigma}$. The fluctuation is defined as:
\begin{equation}
\delta z_i = z_i - \bar{z} = \sum_{j \neq i} (J_{ij} - p_{ij}).
\end{equation}

The correlation function we want to compute is:
\begin{equation}
\overline{\delta z_i \, \delta z_j} = \overline{ \left( \sum_{k \neq i} (J_{ik} - p_{ik}) \right) \left( \sum_{l \neq j} (J_{jl} - p_{jl}) \right) } = \sum_{k \neq i} \sum_{l \neq j} \overline{ (J_{ik} - p_{ik})(J_{jl} - p_{jl}) }.
\end{equation}
(By linearity of expectation). 

In our graphs, different bonds/links are independent. The expectation value $\overline{ (J_{ik} - p_{ik})(J_{jl} - p_{jl}) }$ is zero unless the two bonds are exactly the same bond. Since $J_{ab} = J_{ba}$, the bond is unordered: the pair $\{i,k\}$ must equal the pair $\{j,l\}$.
Consider two separate cases: $i = j$ and $i \neq j$.

\paragraph*{\textbf{Case 1: $i = j$ (same site)}:}
When $i = j$, we have:
\begin{equation}
\overline{\delta z_i^2} = \sum_{k \neq i} \sum_{l \neq i} \overline{ (J_{ik} - p_{ik})(J_{il} - p_{il}) }.
\end{equation}
We split the double sum into two subcases:

\textbf{Subcase (a): $k = l$.} Then the two factors involve the same bond. The expectation value is:
\begin{equation}
\overline{ (J_{ik} - p_{ik})^2 } = \overline{ J_{ik}^2 - 2p_{ik}J_{ik} + p_{ik}^2 } = \overline{ J_{ik}^2}  - 2 \overline{ J_{ik}}p_{ik} + p_{ik}^2 = \overline{ J_{ik}}  - 2 p_{ik}^2 + p_{ik}^2  = p_{ik} -  p_{ik}^2 = p_{ik} (1 -  p_{ik}).
\end{equation}
We have used $J_{ik}^2 = J_{ik}$, and $\overline{J_{ik}} = p_{ik}$.

\textbf{Subcase (b): $k \neq l$.} Then the two bonds $(i, k)$ and $(i, l)$ are independent, so no correlations between them. Therefore,
\begin{equation}
\overline{\delta z_i^2} = \sum_{k \neq i} p_{ik}(1 - p_{ik}) \approx \sum_{k \neq i} p_{ik} = \bar{z}.
\end{equation}
(In the sparse limit $p_{ik} \ll 1$, we have $p_{ik}(1-p_{ik}) \approx p_{ik}$)

\paragraph*{\textbf{Case 2: $i \neq j$ (different sites)}:}
When $i \neq j$, we have:
\begin{equation}
\overline{\delta z_i \delta z_j} = \sum_{k \neq i} \sum_{l \neq j} \overline{ (J_{ik} - p_{ik})(J_{jl} - p_{jl})}.
\end{equation}
The only non-zero contribution in the above summations comes from a single term with $k=j$ and $l=i$, and thus
\begin{equation}
\overline{\delta z_i \delta z_j} = \overline{ (J_{ij} - p_{ij})(J_{ji} - p_{ji})} = \overline{ (J_{ij} - p_{ij})^2 } = p_{ij}(1 - p_{ij}),
\end{equation}
since $J_{ij} = J_{ji}$ and $p_{ij} = p_{ji}$. In the sparse graph limit,
\begin{equation}
\overline{\delta z_i \delta z_j} \approx p_{ij} = \frac{1}{|i-j|^{1+\sigma}} \quad \text{for } i \neq j.
\end{equation}

Combining both cases, we obtain,
\begin{equation}
\boxed{
\overline{\delta z_i \delta z_j} = 
\begin{cases}
\displaystyle \sum_{k \neq i} p_{ik}(1-p_{ik}) \approx \bar{z}, & i = j, \\[1em]
p_{ij}(1-p_{ij}) \approx \dfrac{1}{|i-j|^{1+\sigma}}, & i \neq j.
\end{cases}
}
\label{eq:supp_mass_corr}
\end{equation}

\subsection{Disorder averaging using replica trick}
We treat the emergent mass disorder under Gaussian truncation i.e. keep the first two cumulants of $\delta z_i$ while all higher-order cumulants are discarded: $\overline{\delta z_i} = 0$, $\overline{\delta z_i \delta z_j} = \bar z\,\delta_{ij} + p_{ij}(1-\delta_{ij})$, Eq.\,\eqref{eq:supp_mass_corr}. This is in spirit of the random-mass disorder problems used by Harris and by Weinrib and Halperin~\,\cite{harris1974effect, weinrib1983critical, dotsenko2005introduction}.

Writing $z_i = \bar z + \delta z_i$ in Eq.\,\eqref{eq:supp_mass_terms}, we note that for a single graph realization the mass disorder enters only through a term linear in $\delta z_i$, coupled to the local energy density $|\vect\phi_i|^2$:
\begin{equation}
S_{\rm dis}[\vect\phi;\delta z] = \lambda \sum_i \delta z_i \, |\vect\phi_i|^2, \qquad \lambda \equiv \frac{1}{16\mu^3},
\end{equation} 
while the disorder-independent bare mass $r = \frac12(-1+\frac{1}{2\mu}) + \lambda\bar z$.

Since the bond disorder is quenched, we perform the disorder averaging via the replica trick. The average over the disorder $\delta z_i$ acts only on the quadratic term because $U(\vect{\phi}_{j})$ contains no $\delta z_j$ dependence. Thus, averaging over $m$ copy of replicas,
\begin{equation}
\overline{Z^m} = \int \left[ \prod_{A=1}^m \mathcal{D}\vect{\phi}_A \right] \exp\left( \sum_{A=1}^m \sum_j U(\vect{\phi}_{A,j}) - \sum_A S_0[\vect\phi_A] \right) \overline{ \exp\left( - \lambda \sum_{A=1}^m \sum_i \delta z_j \, |\vect\phi_{A,j}|^2\right) }.
\end{equation}
For a linear functional $X = \sum_{A,i} c_{A,i}\,\delta z_i$ of a zero-mean Gaussian field, $\overline{e^{X}} = e^{\frac{1}{2} \overline{X^2}}$ holds exactly. With $c_{A,i} = -\lambda\,|\vect\phi_{A,i}|^2$,
\begin{equation}
\overline{\exp\left(-\lambda \sum_{A,i}\delta z_i\,|\vect\phi_{A,i}|^2\right)} = \exp\left(\frac{\lambda^2}{2}\sum_{A,B=1}^m \sum_{i,j} \overline{\delta z_i \delta z_j}\;|\vect\phi_{A,i}|^2 |\vect\phi_{B,j}|^2\right).
\end{equation}
Substituting Eq.\,\eqref{eq:supp_mass_corr} in the above equation yields,
\begin{equation}
\exp\left(\frac{\lambda^2 \bar z}{2}\sum_{A,B}\sum_i |\vect\phi_{A,i}|^2 |\vect\phi_{B,i}|^2
\;+\;
\frac{\lambda^2}{2}\sum_{A,B}\sum_{i\neq j} p_{ij}\,|\vect\phi_{A,i}|^2 |\vect\phi_{B,j}|^2
\right)
\end{equation}
The $A=B$ terms in both sums renormalize the bare quartic coupling, $u \to u - \lambda^2\bar z/2$ (short-range contribution), and generate a nonlocal single-replica self-interaction $\propto p_{ij}$ (long-range contribution). The $A\neq B$ terms are the genuine disorder-induced replica-mixing operators. Identifying
\begin{equation}
\Delta_{\rm SR} \equiv \lambda^2 \bar z, \qquad \Delta_{\rm LR} \equiv \lambda^2,
\end{equation}
we have
\begin{equation}
S_v = -\frac{\Delta_{\rm SR}}{2}\sum_{A,B}\sum_i |\vect\phi_{A,i}|^2 |\vect\phi_{B,i}|^2 \;-\; \frac{\Delta_{\rm LR}}{2}\sum_{A,B}\sum_{i\neq j} p_{ij}\,|\vect\phi_{A,i}|^2 |\vect\phi_{B,j}|^2,
\label{eq:Sv_gauss_discrete}
\end{equation}
i.e., a short-range uncorrelated Harris-type term and a LR Weinrib--Halperin-type term.

\subsubsection{The propagator}
The disorder-averaged action is translationally invariant. The propagator $\vect{G}_{ij}$ satisfies:
\begin{equation}
\sum_{k} \vect{G}^{-1}_{ik} \vect{G}_{kj} = \delta_{ij},
\end{equation}
which yields in Fourier space 
\begin{equation}
\tilde{\vect{G}} (q)  = \left[ 2 \mu + \tilde{p}(q) \right]^{-1}.
\end{equation}
In discrete Fourier space,
\begin{equation}
\tilde{p}(q) = \sum_{j \neq i} \frac{e^{-iq(x_i - x_j)}}{|i-j|^{1+\sigma}} = \underbrace{\sum_{j \neq i} \frac{1}{|i-j|^{1+\sigma}}}_{\equiv \bar{z}} \;+\; \sum_{j \neq i} \frac{e^{-iq(x_i - x_j)} - 1}{|i-j|^{1+\sigma}}.
\end{equation}

For $0 < \sigma < 1$ and small $q$ (but $q \gg 1/N$), the second term yields the leading non-analytic contribution:
\begin{equation}
\tilde{p}(q) = \bar{z} + c_2 |q|^{\sigma} + \cdots , \qquad q \to 0,
\end{equation}
where $c_2$ is a positive constant. Note that average degree $\bar{z}$ is finite for $\sigma > 0$. Substituting into the propagator gives:
\begin{equation}
\tilde{\vect{G}}(q) = \frac{1}{2\mu + \bar{z} + c_2 |q|^{\sigma} + \cdots} \approx \frac{1}{2\mu + \bar{z}} - \frac{c_2}{(2\mu + \bar{z})^2} |q|^{\sigma} + \cdots
\end{equation}
The constant term contributes to the mass, while the $|q|^{\sigma}$ term is the leading long-range kinetic contribution. For $\sigma > 1$, the $|q|^{\sigma}$ term is replaced by $q^2$ (short-range kinetics).

\subsubsection{Continuum limit}

The propagator in Fourier space,
\begin{equation}
\tilde{G}(q) = \frac{1}{m^2 + c_2 |q|^{\sigma}}, \qquad 0 < \sigma < 1,
\end{equation}
where $m^2 = 2\mu + \bar{z}$ is the renormalized mass. In the real space:
\begin{equation}
G(x) = \int_{-\infty}^{\infty} \frac{dq}{2\pi} \, \frac{e^{iqx}}{m^2 + c_2 |q|^{\sigma}}.
\end{equation}
Since the integrand is even in $q$, this becomes
\begin{equation}
G(x)=\frac{1}{\pi}\int_0^\infty dq\,
\frac{\cos(qx)}{m^2+c_2q^\sigma}.
\end{equation}

To determine the asymptotic behavior at large $|x|$, we note that the dominant contribution arises from the small-$q$ region. Expanding the denominator for small $q$ gives
\begin{equation}
\frac{1}{m^2+c_2q^\sigma}
=
\frac{1}{m^2}
\frac{1}{1+\frac{c_2}{m^2}q^\sigma} = \frac{1}{m^2}
-\frac{c_2}{m^4}q^\sigma
+O(q^{2\sigma})..
\end{equation}

Substituting into the Fourier integral,
\begin{equation}
G(x)
=
\frac{1}{\pi m^2}
\int_0^\infty dq\,\cos(qx)
-\frac{c_2}{\pi m^4}
\int_0^\infty dq\, q^\sigma \cos(qx)
+\cdots
\end{equation}
The first term contributes to the mass term, as discussed, and does not affect the long-distance behavior. Thus, the leading asymptotic contribution is
\begin{equation}
G(x)\sim
-\frac{c_2}{\pi m^4}
\int_0^\infty dq\, q^\sigma \cos(qx).
\end{equation}

Now, use the standard integral
\begin{equation}
\int_0^\infty dq\, q^\sigma \cos(qx) = \frac{\Gamma(1+\sigma) \cos\left[\frac{\pi(1+\sigma)}{2}\right]} {|x|^{1+\sigma}}.
\end{equation}
and the identity
\begin{equation}
\cos\left[\frac{\pi(1+\sigma)}{2}\right]
=
-\sin\left(\frac{\pi\sigma}{2}\right),
\end{equation}
to find
\begin{equation}
\int_0^\infty dq\, q^\sigma \cos(qx)
=
-\frac{
\Gamma(1+\sigma)
\sin\left(\frac{\pi\sigma}{2}\right)}
{|x|^{1+\sigma}}.
\end{equation}

Therefore, the asymptotic decay of the kinetic term of the propagator
\begin{equation}
G(x)\sim \frac{A}{|x|^{1+\sigma}},
\end{equation}
with amplitude
\begin{equation}
A=
\frac{
c_2\,\Gamma(1+\sigma)
\sin\left(\frac{\pi\sigma}{2}\right)}
{\pi m^4}.
\end{equation}

In the continuum limit, the effective action becomes
\begin{align}
S_{\rm eff}[\vect\phi] = \;&\frac{1}{2}\sum_{A=1}^m \int dx\,dy\, \frac{\vect\phi_A(x)\cdot\vect\phi_A(y)}{|x-y|^{1+\sigma}} \;+\; r\,\sum_{A=1}^m \int dx\, |\vect\phi_A(x)|^2 \;+\; u_{\rm eff}\,\sum_{A=1}^m \int dx\, |\vect\phi_A(x)|^4 \nonumber\\
&-\; \frac{\Delta_{\rm LR}}{2}\sum_{A=1}^m \int dx\,dy\, \frac{|\vect\phi_A(x)|^2 |\vect\phi_A(y)|^2}{|x-y|^{1+\sigma}} \;-\; \frac{\Delta_{\rm SR}}{2}\sum_{A,B=1}^m{ }' \int dx\, |\vect\phi_A(x)|^2 |\vect\phi_B(x)|^2 \nonumber\\
&-\; \frac{\Delta_{\rm LR}}{2}\sum_{A,B=1}^m{}' \int dx\,dy\, \frac{|\vect\phi_A(x)|^2 |\vect\phi_B(y)|^2}{|x-y|^{1+\sigma}},
\label{eq:Seff_gauss_continuum}
\end{align}
where $\sum'_{A,B}$ denotes $A\neq B$ only, $u_{\rm eff}=u-\Delta_{\rm SR}/2$, and $r$ is the disorder-averaged bare mass identified earlier. The disorder sector $S_v$,
\begin{equation}
S_v \equiv -\frac{\Delta_{\rm SR}}{2}\sum_{A,B}{}' \int dx\, |\vect\phi_A(x)|^2 |\vect\phi_B(x)|^2 \;-\; \frac{\Delta_{\rm LR}}{2}\sum_{A,B}{}' \int dx\,dy\,\frac{|\vect\phi_A(x)|^2 |\vect\phi_B(y)|^2}{|x-y|^{1+\sigma}}.
\label{eq:Sv_gauss_continuum}
\end{equation}
Both the long-range disorder vertex and the long-range kinetic term thus controlled by the same parameter $\sigma$.

\section*{Conventional SR uncorrelated  and LR correlated disorder}
\label{sup:HWH_criteria}
The conventional setup involves a ``clean'' critical system in $d$ spatial dimensions described by an action $S_0$, with a quenched random disorder field $h(x)$ coupled to a single scalar operator $\mathcal{O}_0 (x)$: $S = S_0 + \int d^dx\, h(x)  \mathcal{O}_{0}(x)$. For SR uncorrelated disorder:  $\overline{ h(x)} = 0, \overline{ h(x) h(y)} = v_{\rm{SR}} \, \delta^{(d)}(x-y)$, the effective replicated action is
\begin{equation}
S_{\rm{rep}} =  \sum_{A} S_{0, A} -  \frac{v_{\rm{SR}}}{2}  \sum_{A,B}{}' \int d^dx\, \mathcal{O}_{0, A}(x)  \mathcal{O}_{0, B}(x),
\label{eq:action_Harris}
\end{equation}
where $A, B$ are replica indices. Next, we consider the LR correlated disorder $\overline{ h(x) h(y)} =v_{\rm{LR}} \, \, |x-y|^{-a}$, which yields a non-local disorder-induced interaction in the replicated action,
\begin{equation}
S_{v} \sim - \frac{v_{\rm{LR}}}{2} \sum_{A,B}{}' \int d^d x \, d^d y \,\, \frac{\mathcal{O}_{0, A} (x) \mathcal{O}_{0, B}(y)}{|x-y|^a}.
\label{eq:action_WH}
\end{equation}

In the standard Harris and WH analyses, the clean action $S_0$ is that of $|\vect{\phi}|^4$ theory describing criticality at its Wilson-Fisher fixed point characterized by the correlation-length exponent $\nu$ and the operator is taken to be the local energy density, $\mathcal{O}_0 (x) = \vect{\phi}^2 (x)$, representing random mass disorder. A disordered theory may flow to a `clean' fixed point or to a `random' fixed point determined by the beta function $\beta_v$. However, the question of disorder \emph{relevance} can already be addressed by the scaling dimension of disorder-induced operator around the clean fixed point. 

The scaling dimension of the local operator at the clean fixed point $[\mathcal{O}_0] = d - 1/\nu$\,\cite{cardy1996scaling}, which determines the scaling dimension of the disorder-induced interactions
\begin{equation}
[v_{\rm{SR}}] = 2/\nu - d,\quad [v_{\rm{LR}}] =  2/\nu - a.
\end{equation}
SR disorder is irrelevant if $[v_{\rm{SR}}] < 0$, recovering the Harris criterion $\nu > 2/d$. Similarly, for LR correlated disorder, the disorder irrelevance criterion $\nu > 2/a$ for $a<d$. For $a>d$, one can argue from the integrability of the LR disorder kernel that it effectively becomes short-range under coarse-graining and would eventually follow the Harris criterion\,\cite{harris1974effect, weinrib1983critical}.

\end{document}